\documentclass[preprint,12pt,superscriptaddress,floatfix,pra]{revtex4-1}
\usepackage{graphicx}
\usepackage{hyperref}
\usepackage{setspace}

\renewcommand{\onlinecite}[1]{\cite{#1}}

\usepackage{amssymb}
\usepackage{amsmath}
\usepackage[title,titletoc]{appendix}

\newcommand{\U}{\mathrm{U}}

\newcommand{\eqnref}[1]{Eq.~(\ref{#1})}

\usepackage{xcolor}

\newcommand{\Yb}{\ensuremath{^{171}\mathrm{Yb}^+}}
\newcommand{\bra}[1]{\left \langle #1 \right |}
\newcommand{\ket}[1]{\left | #1 \right \rangle}

\newcommand{\vect}[1]{\overrightarrow{#1}}

\begin{document}

\title{Discrete Time Crystals}

\author{Dominic V. Else}
\affiliation{Department of Physics, Massachusetts Institute of Technology, MA 02139, USA}

\author{Christopher Monroe}
\affiliation{Joint Quantum Institute, Center for Quantum Information and Computer Science}
\affiliation{Department of Physics, University of Maryland, College Park, MD 20742, USA}

\author{Chetan Nayak}
\affiliation{Station Q, Microsoft Research, Santa Barbara, California 93106-6105, USA}
\affiliation{Department of Physics, University of California,  Santa Barbara, California 93106, USA}

\author{Norman Y. Yao}
\affiliation{Department of Physics, University of California, Berkeley, CA 94720, USA}
\affiliation{Materials Sciences Division, Lawrence Berkeley National Laboratory, Berkeley CA 94720, USA}


\begin{abstract}
Experimental advances have allowed for the exploration of nearly isolated quantum many-body systems whose coupling to an external bath is very weak. A particularly interesting class of such systems is those which do not thermalize under their own isolated quantum dynamics. In this review, we highlight the possibility for such systems to exhibit new \emph{non-equilibrium} phases of matter. In particular, we focus on ``discrete time crystals'', which are many-body phases of matter characterized by a spontaneously broken discrete time translation symmetry. We give a definition of discrete time crystals from several points of view, emphasizing that they are a non-equilibrium phenomenon, which is stabilized by many-body interactions, with no analog in non-interacting systems. We explain the theory behind several proposed models of discrete time crystals, and compare a number of recent realizations, in different experimental contexts.
\end{abstract}
\maketitle

\tableofcontents

\section{Introduction}
\label{sec:Intro}

Much of the mystery and magic of quantum mechanics remains concealed if we
focus solely on systems in thermal equilibrium. 
Indeed, many applications, such as quantum metrology,  computing, and  communication rely upon the production, preservation, and manipulation of systems far from equilibrium; their efficacy depends on evading or at least impeding thermalization, which naturally leads to the loss of locally stored quantum information.
In recent years, advances in the control of highly-isolated quantum systems have enabled experiments to probe multiple facets of thermalization and of its failure \cite{Nandkishore15, Gring12, BlochMBL}.
Hence, it has become possible to play with interesting non-equilibrium quantum states of matter that do not follow the usual tenets of thermal statistical mechanics.
In this review, we focus on a particular class: \emph{time crystals} in periodically-driven (Floquet), isolated quantum systems.
As will become apparent, although our focus is on time crystals, the viewpoint of this review is that they serve as a paradigmatic example of a much broader class of physical phenomena, namely, non-equilibrium phases of matter.

No system is completely isolated, and no drive is perfectly periodic. However, by studying the ideal limit, one can understand the dynamics that governs the evolution of a real system until the time scale at which noise in the drive and the coupling to the environment take over.
In this ideal limit,
a Floquet system is characterized by a Hamiltonian that is periodic in time with period $T$: $H(t) = H(t+T)$.
We can make sharp distinctions between different Floquet phases of matter when the observables of such a system, measured at stroboscopic times $t=kT$, with $k\in\mathbb{Z}$, settle into a steady state.
A discrete time crystal is a distinct Floquet phase of matter that spontaneously breaks the discrete time-translational symmetry $t\rightarrow t + T$ down to $t\rightarrow t + nT$ for some integer $n>1$.
It can be probed in experiments by measuring an order parameter that transforms under the symmetry.

There is an entire universe of time-dependent non-equilibrium phenomena, but the term \emph{spontaneous symmetry-breaking} is reserved for a particular set of properties that are normally associated with equilibrium phases of matter. Indeed, we usually think of spontaneous symmetry breaking as a property of a ground state or a thermal ensemble,
and, in fact, time crystals were originally envisioned 
by Wilczek \cite{Wilczek12} and Shapere and Wilczek \cite{Shapere12} as an equilibrium state of matter in which continuous time-translation symmetry is broken.
However, subsequent work \cite{Bruno13a,Bruno13b,Bruno13c,Nozieres13} showed that an equilibrium time crystal is not possible, culminating in the no-go theorem of Watanabe and Oshikawa \cite{Watanabe15}.
Hence, time crystals can only exist in systems that are not in thermal equilibrium, such as the Floquet systems that form the focus of this review.
But, in such a case, it might be tempting to
classify a number of superficially similar-looking non-equilibrium phenomena (e.g.~period-doubling bifurcations, Faraday waves, etc) as time crystals \cite{Yao18b}. 
However, such time-dependent phenomena generally
cannot be classified as \emph{phases of matter}
and share few similarities with equilibrium order.
As we shall see, a discrete time crystal is a
spontaneous symmetry-breaking phase of
matter according to a definition that is
a natural generalization of the equilibrium notion \cite{Else16b,Else17a, Khemani16, Yao17}.
Consequently, the time crystals discussed in this review distinctly retain the original spirit of spontaneously broken time-translational symmetry. In this review, we primarily consider discrete time-translational symmetry in periodic systems, but we also briefly touch on contiuous time-translational symmetry in undriven systems.
Remarkably, many of the key properties of equilibrium ordered states emerge in discrete time crystals, which can occur due to the nature of the far-from-equilibrium steady states in closed, driven quantum systems.

Even a completely isolated quantum system
will generically thermalize.
In particular, subsystems of the full quantum system can act as heat baths for each other, and the expectation values of \emph{local} observables at long times, evolving under a static Hamiltonian, will look just like those of a thermal state in the canonical ensemble of the same Hamiltonian. This process is known as \emph{thermalization}.
In driven systems, energy is not conserved since the system can absorb energy from the drive via Floquet heating. Thermalization in a periodically-driven system is the approach, through this heating process, to a state that is locally indistinguishable from an infinite temperature state \cite{DAlessio14,Ponte15a,Lazarides_1403}.
While integrable systems such as non-interacting systems have long been known to evade thermalization, they require fine-tuning and will still thermalize in the presence of even weak perturbations.
One starting point for the recent exploration of non-equilibrium Floquet phases has been the revelation that there are more generic classes of systems whose failure to thermalize \emph{does not} require fine-tuning.

In a disordered system, some of the parameters in the Hamiltonian vary randomly as a function of space.
When disorder is sufficiently strong, an isolated one-dimensional quantum system will
undergo a phenomenon called
\emph{many-body localization} (MBL) ~\cite{Basko06a,Basko06b,Oganesyan07,Znidaric08,Pal10,Bardarson12,Serbyn13a,Serbyn13b,Bauer13,Huse14,Nandkishore15}.
In the MBL phase, there are extensively many emergent quasi-local conserved quantities, and as a consequence, the system undergoing isolated quantum evolution retains memory of its initial state forever instead of thermalizing. Crucially, seminal recent work has demonstrated that MBL can also occur in periodically-driven systems
\cite{Abanin14,Ponte15a,Ponte15b,Lazarides15,Iadecola15,
Khemani16,vonKeyserlingk16a,Else16a,
Potter16,Roy16,vonKeyserlingk16b}, where it prevents such a Floquet system from heating to infinite temperature.


Although the MBL regime has a convenient theoretical definition, it is difficult to achieve in practice since it requires strong disorder and vanishing coupling to the environment. 
However, there is another regime in which non-trivial Floquet states of matter can be realized. 
When a system is periodically-driven at  high frequency, thermalization can occur very slowly.
The intuition is as follows: since the system can only absorb energy from the drive in multiples of the drive frequency $\Omega=2\pi/T$, it must make many local rearrangements in order to do so, a process whose amplitude is exponentially small in the ratio of $\Omega$ to the local bandwidth, $\Omega_0$. 
Thus, the system takes an exponentially-long time to reach the infinite-temperature state. During this time interval, the system settles into an exponentially-long lived non-equilibrium steady state, a process
called \emph{Floquet prethermalization} \cite{Kuwahara16,Abanin17}.
Non-equilibrium quantum steady-states can 
emerge in the \emph{prethermal} regime that are
strictly forbidden in thermal equilibrium.


The aforementioned strategies focus on attempting to stabilize Floquet phases for either infinitely or exponentially long time-scales, however, one of the key lessons learned from recent experiments \cite{ZhangJ17,ChoiS17,Rovny18a,PalS18,OSullivan18} is that the signatures of time crystalline order might occur more readily in nature than expected.
Indeed, these signatures can arise during the transient early stages of Floquet heating, before thermalization or even prethermalization occurs.
To this end, experimental platforms where the initial stages of Floquet heating can be slowed down, for example by using a combination of weak disorder (i.e.~not strong enough to realize MBL) and an interplay between long-range, power-law interactions and dimensionality, may constitute another promising set of venues for exploring time crystalline order \cite{Ho17,Kucsko18}. 

Let us discuss how one can directly realize and observe phenomena associated with many-body localization and prethermalization. To do this one generally has to go beyond the usual setting of solid-state physics, i.e.\ electrons in a conventional solid, because the electron-phonon coupling usually causes rapid thermalization.
However, careful engineering can give rise to degrees of freedom that are better isolated from their environment. Often, a motivation for developing such techniques has been the goal of realizing quantum computation, since the qubits in a quantum computer must be very well isolated; in this review, we instead focus on applications to the study of many-body quantum dynamics.
Deliberately-designed nano- and micro-structures \cite{devoret2004superconducting,clarke2008superconducting,barends2016digitized} or isolated defects in the solid-state \cite{koppens2006driven,schirhagl2014nitrogen, doherty2013nitrogen, koehl2011room} enjoy some level of isolation from the environment. 
As a result, these degrees of freedom, which are sometimes also called ``artificial atoms", do not thermalize as fast as electrons in the solid state normally do. Nuclear spins in solids offer even better isolation, and can be manipulated using the mature tools of nuclear magnetic resonance \cite{harris1986nuclear,callaghan1991principles,vandersypen2001experimental}.

The cleanest platforms for the study of quantum non-equilibrium phenomena are ``artificial solids," comprised of trapped ions or neutral atoms, where coupling to a thermal bath can be exquisitely controlled. 
It has proven difficult to use artificial solids to emulate real solids because, for instance, the lack of a phonon bath makes it onerous to directly cool such a system into a non-trivial quantum ground state.
However, they offer the ideal platform for observing the dynamics of isolated quantum systems driven far from equilibrium!


\section{Phases of matter out of equilibrium}
\label{sec:phases_ooe}

How should we define phases of matter in isolated quantum systems out of equilibrium? Ideally, we would find a way to extend the notion of phases of matter in thermal equilibrium that allows us to consider new and interesting phenomena without being so general that it becomes meaningless.
In this section, we will outline several different approaches. In Section \ref{subsec:manybodysteady}, we introduce the idea of a ``many-body steady state'', which generalizes a thermal equilibrium state and whose distinct regimes we could call ``phases of matter''. We will not attempt a precise definition of a many-body steady state, in order not to unduly rule out potential yet-to-be-discovered phenomena. In Sections \ref{sec:locprotected} and \ref{subsec:crypto} we consider potentially more restrictive scenarios, where the connection between out-of-equilibrium phases and equilibrium phases can be made very explicit: in Section \ref{sec:locprotected} in terms of eigenstates of the Hamiltonian $H$ or Floquet evolution operator $U_{\text{F}}$; and in Section \ref{subsec:crypto} in terms of ``crypto-equilibrium'', which is approximate thermal equilibrium in a rotating frame.

\subsection{Many-body steady states}
\label{subsec:manybodysteady}

The notion of a \emph{phase of matter} is one that is very well established in thermal equilibrium. The thermal state of the system, given by the Gibbs state
\begin{equation}
\label{gibbs_state}
\rho_{\mathrm{Gibbs}} = \frac{1}{Z} e^{-\beta H}
\end{equation}
can exhibit sharply distinct regimes, and we call these phases of matter. As mentioned in Sec. \ref{sec:Intro}, as a result of thermalization, the Gibbs state correctly describes the late time behavior of local observables in an isolated quantum system. In a periodically-driven system, thermalization leads to \eqnref{gibbs_state} with $\beta=0$, so there are no non-trivial phases of matter.

When thermalization does not occur, for example in systems that exhibit MBL, at late times we might still expect the expectation values of local observables to relax to those of a steady state, but this state will \emph{not} be described by a Gibbs state, but instead be some non-thermal state (we will include in the term ``steady state'' cases where the local observables oscillate with some period). We will be interested in such late time steady states to the extent that they are \emph{not fine-tuned} -- that is, the properties of the steady state does not depend on some fine-tuned parameter of the Hamiltonian.
In particular, we want to consider steady states whose properties are robust to, or even stabilized by, many-body interactions.
 We also want to consider properties that become sharply defined in the thermodynamic limit as the system size goes to infinity. We refer to a steady state with such robust properties as a \emph{many-body steady state}. A many-body steady state is a natural generalization of a thermal equilibrium state (Gibbs state), and this will allow much of the phenomenology of phases of matter to be carried over from thermal equilibrium to many-body steady states. In particular the notion of spontaneous symmetry breaking in many-body steady states will be discussed further in Section \ref{sec:TTSB}.
In a periodically-driven system, with $H(t+T) = H(t)$,
we are interested in systems in which the expectation values of local observables at stroboscopic times $t=kT$, with $k\in \mathbb{Z}$, relax to a many-body steady state. If discrete time-translation symmetry is broken, we weaken this requirement to relaxation to a many-body steady state at times $t=nkT$ for some integer $n$.

\subsection{Localization-protected quantum order in MBL}
\label{sec:locprotected}
The connection between the non-equilibrium phases of matter we consider in this review and traditional equilibrium phases can be made even stronger, at the cost of introducing less experimentally observable considerations. We will do this by considering eigenstates of the time evolution operator.

First of all, we recall that any Hamiltonian $H$ whose isolated evolution is thermalizing, as defined in the previous section, is believed to satisfy the \emph{eigenstate thermalization hypothesis} (ETH)  \cite{Deutsch91,Srednicki94,Rigol08,DAlessio15}, which posits that any eigenstate of $H$, with eigenvalue $E$, looks identical to a Gibbs state $\rho = \frac{1}{Z} e^{-\beta H}$ whose expectation value of the energy is $E$, on any subsystem small compared with the total system size.

There is also a variant of ETH that applies to perodically-driven isolated systems \cite{DAlessio14,Ponte15a,Lazarides_1403}. For a system that heats to infinite temperature, as discussed in the introduction, the eigenstates of the Floquet operator $U_\text{F}$ look identical to the infinite temperature state on subsystems small compared to the system size. Here the Floquet operator $U_{\text{F}}$ is the unitary operator that generates the time evolution over one time cycle: it can be defined as
\begin{equation}
\label{eq:floquetop}
U_\text{F} = \mathcal{T} \exp\left(-i\int_0^T H(t) dt \right),
\end{equation}
where $H(t)$ is the time-periodic Hamiltonian that generates the dynamics, and $\mathcal{T}$ is the time-ordering symbol.

In this review, we are interested in systems that are not thermalizing, hence are not expected to obey ETH either. The question is, what is the nature of the eigenstates in that case. For MBL systems, the answer is remarkable: the eigenstates have all of the properties typically associated with \emph{gapped ground states} of quantum systems \cite{Serbyn13a,Bauer13,Huse14}; in fact, for any eigenstate of an MBL system, one can construct a fictitious quasi-local Hamiltonian for which the eigenstate is the unique gapped ground state. One consequence of this is that the eigenstates of an MBL system have an entanglement entropy on subsystems that scales with the size of the boundary of the subsystem (``area law'') , whereas eigenstates of a Hamiltonian that obeys ETH have entanglement entropy that agrees with the thermodynamic entropy, and consequently scales with the volume of the subsystem (``volume law'').

Because of this property of MBL systems, they can host ``eigenstate phases of matter'' \cite{Huse_1304,Bauer13,Bahri_1307}: in other words, the same notion of phases and phase transitions that exists for gapped quantum systems at zero temperature (which are completely characterized by their ground state) can be applied to the eigenstates of an MBL system. Of course, eigenstates are not themselves directly observable in practice. Nevertheless, when eigenstates display non-trivial phases of matter, in practice these occur simultaneously with more observable properties such as features of the long-time dynamics.


\subsection{Floquet prethermalization and crypto-equilbrium}
\label{subsec:crypto}

Non-trivial Floquet phases of matter can exist in the absence of MBL -- if, for instance,
disorder is weak or even absent -- provided
we slightly relax our notion of a phase of matter to include states that are exponentially but not infinitely long-lived.
If the frequency of the drive is very large compared to the local energy scales of the system, then the system can only
absorb energy from the drive by spreading it out over many excitations. Consequently, heating occurs very slowly \cite{Abanin_1507,Kuwahara_1508,Mori_1509,Abanin17,Bukov_1507,Canovi_1507,
Bukov_1512,Machado_1708},
and there is a long-lived quasi-steady state -- called a ``prethermal'' state -- in which non-trivial states of matter can occur.
Abanin et al. \cite{Abanin17} proved a theorem showing that, in the high-frequency limit (which means that the frequency is much larger than the local energy bandwidth of the system, $\Omega_0$), then there exists a rotating frame transformation, described by a time-dependent unitary $P(t)$, such that the Hamiltonian in the rotating frame, which can be written as
\begin{equation}
H_{\mathrm{rot}}(t) = P^{\dagger}(t) H(t) P(t) - P^{\dagger}(t) \partial_t P(t),
\end{equation}
is approximately time-independent. That is, we can approximate
\begin{equation}
H_{\mathrm{rot}}(t) = H_F + O(e^{-\Omega/\Omega_0}),
\end{equation}
for some time-independent quasi-local Hamiltonian $H_F$.


Consequently, for times short compared to the exponentially-long time $t_\text{Th} \sim e^{\Omega/{\Omega_0}}$, the system appears (in the rotating frame) to be evolving under the time-independent Hamiltonian $H_F$. In particular, assuming that $H_F$ is thermalizing, the system will evolve into a state that locally looks like a Gibbs state $\frac{1}{Z} e^{-\beta H_F}$. We call this process \emph{prethermalizaton}, and the resulting state $\frac{1}{Z} e^{-\beta H_F}$ is the \emph{prethermal state}. In other words, in the rotating frame, the system is effectively in equilibrium. We call this property ``crypto-equilibrium'' to indicate that we can carry over all the traditional notions of equilibrium phases of matter.

It might not be immediately obvious how one can realize \emph{new} phases of matter in crypto-equilibrium, since the evolution is governed by a static Hamiltonian. Of course, to determine the actual evolution of the system, we have to undo the rotating frame transformation. But in the case considered by Ref.~\cite{Abanin17}, the rotating frame transformation vanishes at stroboscopic times $t=kT$, so if we only observe the system at stroboscopic times $t = kT$, then the system appears to be evolving under $H_F$. In other words, we can write the Floquet evolution operator as
\begin{equation}
U_{\text{F}} = e^{-iH_F T} + O(e^{-\Omega/\Omega_0}),
\end{equation}
where $H_F$ is a local time-independent effective Hamiltonian.
Moreover, the rotating frame transformation is ``small'', in the sense that its effect on local observables is of order $\Omega_0 / \Omega \ll 1$. When these properties are violated, as they will be in the case we consider in the next paragraph, the dynamics of the original Hamiltonian, after undoing the rotating frame transformation, can still exhibit sharply distinct non-equilibrium regimes (which nevertheless can be formally characterized through properties of the effective Hamiltonian $H_F$, as we shall see).

In the limit considered in the previous paragraph, the Hamiltonian is equal to a sum of local terms, each of which has an operator norm $\sim \Omega_0$; consequently, the Floquet operator $U_\text{F}$ generating the time evolution over one time period only enacts a tiny rotatation of a state, as far as any local operator is concerned. Suppose, instead, that we require the weaker condition that, for some integer $N$,
$(U_\text{F})^N$ only effects a small rotation of any local operator, or to put it another way, $U_F \approx X$ for some unitary $X$ that satisfies $X^N = 1$. Then we can apply the preceding result with $T \to NT$, and we obtain
\begin{equation}
\label{Nfloquet}
(U_\text{F})^N = e^{i H_\text{F} NT} + O(e^{-\Omega/{N\Omega_0}}).
\end{equation}
In fact, one can prove a stronger result \cite{Else17a}. Suppose the Hamiltonian can be written in the form
$H(t) = {H_0}(t) + V(t)$ where $\Omega_0$, the local bandwidth of $V(t)$, is assumed small compared to $\Omega$.
Meanwhile, ${H_0}(t)$ is not assumed small, but we require that $X^N=1$, where $X={\cal T} \exp(-i\int_0^T {H_0}(t))$.
Then, there exists a (time-independent) unitary transformation ${\cal U}$ such that:
\begin{equation}
\label{prethermal-Floquet}
{\cal U}\, U_F\, {\cal U}^\dagger = X e^{iD T} + O(e^{-\Omega/{N\Omega_0}})
\end{equation}
where $D$ is a local time-independent Hamiltonian satisfying $[D,X]=0$. \eqnref{prethermal-Floquet} indeed implies \eqnref{Nfloquet} with $H_F = \mathcal{U}^{\dagger} D \mathcal{U}$, but \eqnref{prethermal-Floquet} reveals the key property that $H_F$ has a hidden $\mathbb{Z}_N$ symmetry generated by $\mathcal{U}^{\dagger} X \mathcal{U}$. This symmetry is of non-equilibrium origin; we can think of it as a shadow of the discrete time-translation symmetry in the original driving Hamiltonian. Moreover, if $H_F$ realizes a phase spontaneously breaking $X$ or topological phases protected by $X$, this gives rise to non-trivial micromotions in the original driving Hamiltonian. We will see in more detail how this works for prethermal discrete time crystals (which corresponds to the hidden emergent approximate $\mathbb{Z}_N$ symmetry being spontaneously broken) in Section \ref{sec:prethermal_dtc}.

 Finally, we emphasize that the eigenstate properties discussed in the previous section also have analogs in the context of prethermalization. Specifically, if we let $U_\text{F}^{\mathrm{approx}}$ be the approximate Floquet evolution operator corresponding to removing the exponentially small $O(e^{-\Omega/\Omega_0})$ terms in \eqnref{prethermal-Floquet}, then the eigenstates of $U_\text{F}$ will correspond to the eigenstates of $H_F$. Assuming that $H_F$ obeys the ETH, then these eigenstates will look thermal in small subsystems. So whereas in MBL systems, the eigenstates of $U_\text{F}$ look like gapped ground states, allowing us to apply concepts of zero-temperature phases, here the eigenstates look like finite-temperature thermal states, and thus we can apply concepts of finite-temperature phases.

 \section{Prethermal continuous time crystals}
 \label{sec:prethermal-continuous}
 
 In the rest of this review, we mainly talk about periodically driven systems, but in this section we briefly want to mention time crystals in undriven, energy-conserving systems. Here, we can also exploit a form of prethermalization. When this occurs, it can give rise to a time crystal that breaks \emph{continuous} time translation symmetry (row 3, Figure 3) \cite{Urbina82,Autti18,Kreil18}. We will discuss a simple model of this in the present subsection and focus on discrete time crystals in the rest of this review.
Consider a three dimensional spin-1/2 system in a large magnetic field governed by  Hamiltonian,
\begin{equation}
\label{eqn:XXZ}
  H = -h^z \sum_i S_i^z - h^x \sum_i S_i^x - \sum_{\langle i,j \rangle}\left[ J^x S_i^{x} S_j^{x} + J^y S_i^{y} S_j^{y} + J^z S_i^{z} S_j^{z}\right],
\end{equation}
where the sum over $\langle i,j \rangle$ is over nearest neighbor sites, and $h^z$ is much larger than all the other couplings. As is well known, one can remove the effect of the large magnetic field $h^z$ by moving to a rotating frame, and then to zero-th order in $1/h^z$ we can ignore rapidly oscillating terms in the Hamiltonian in the rotating frame, which gives an effective Hamiltonian
\begin{equation}
D_0 = \sum_{\langle i, j \rangle} \left[J (S_i^x S_j^x + S_i^y S_j^y) + J^z S_i^z S_j^z \right], \quad J = \frac{J_x + J_y}{2}
\end{equation}
Observe that this Hamiltonian in fact has a $\U(1)$ symmetry generated by $S^z$. What is less well known is that this ``hidden'' symmetry is actually present at higher orders in $1/h^z$ as well. In fact, one can construct \cite{Abanin17} a static local unitary rotation $\mathcal{U}$ such that
\begin{equation}
\label{eq:static-pretherm-Ham}
\mathcal{U} H \mathcal{U}^{\dagger} = -h^z \sum_i S_i^z + D + O( e^{-h^z/\lambda}),
\end{equation}
where $\lambda = \max\{ |J^x|, |h^x|, |J^y|, |J^z| \}$, and
where \begin{equation}
D = D_0 + O(h^z/\lambda),
\end{equation}
and the higher order terms also preserve the $\mathrm{U}(1)$ symmetry generated by $S^z$.

This is the static analog of the theorem \cite{Kuwahara16,Abanin17} governing prethermalization in Floquet systems. Hence the same notation $D$: this is the effective Hamiltonian that governs the dynamics of the system until exponentially-late times. The reason that thermalization is so slow is that many ``spin waves'' must be created, each with energy $\sim\lambda$, in order for the system to explore states with different values of $\sum_i S_i^z$; such processes are very high order in $h^x$.
For times less than $t_* \sim e^{{h^z}/\lambda}$, we can ignore the $\mathrm{U}(1)$ breaking terms in \eqnref{eq:static-pretherm-Ham}.

The Hamiltonian $D_0$ (and therefore, presumably, its perturbed version $D$) exhibits an XY ferromagnetic phase, which is the magnetic analog of a superfluid phase, at low enough temperatures.
Thus, if we start in an initial state that is low energy with respect to $D$, we expect it to prethermalize to a superfluid state with order parameter $\langle S_i^{+} \rangle \neq 0$, where $S_i^{+} = S_x + i S_y$. In a superfluid, the order parameter rotates at a frequency set by the chemical potential $\mu$ with respect to the full Hamiltonian $\mathcal{U} H \mathcal{U}^{\dagger}$ (which is determined by the energy and $\langle S^z \rangle$ of the initial state). At very late times $\gtrsim t_*$, the $\U(1)$ breaking terms allow the chemical potential to relax to zero, and the rotations cease.

Another way to say this is that for an initial state with low energy with respect to $D$, the prethermal state has relatively low entropy (as measured by the subsystem entanglement entropy), but such an initial state is a very high energy state with respect to the full Hamiltonian $H$ in \eqnref{eqn:XXZ} since the nonzero $\langle S_i^{+} \rangle$ means that the spins are not fully aligned with the large magnetic field. Hence, the system will eventually thermalize into a high-temperature state with respect to $H$ (with high entropy), but this takes an exponentially long time $t_* \sim e^{{h^z}/\lambda}$ due to the approximate symmetry generated by $\mathcal{U}^{\dagger} \sum_i S_i^z \, \mathcal{U}$. As we shall see, this is analogous to the prethermal discrete time crystal, which is an exponentially-long lived low-temperature state of a static Hamiltonian $D$ that eventually evolves into the infinite-temperature state.

This scenario is not ``fine-tuned'': we only need to increase $h^z$ linearly in order to increase the lifetime of the time crystal exponentially.
Note, however, that we have assumed that the system is completely isolated. If the system is not isolated, then
the periodic rotation of the order parameter will cause the system to emit radiation, and this radiation will
cause the system to decay to its true ground state \cite{Bruno13a,Bruno13c}. It is no longer necessary to create many spin waves to change the value of $\sum_i S_i^z$ since the energy can, instead, be carried away by photons or phonons.
 
The Hamiltonian in \eqnref{eqn:XXZ} is rather general and can be realized in a variety of systems.
In the NMR experiment of Ref.~\cite{Urbina82},
the $^{19}$F nuclear spins in CaF$_2$ interact via dipolar interactions, through which they order, and are subject to a large magnetic field $h^z$, which causes them to rotate, as in \eqnref{eqn:XXZ}. 
The magnetic field does not appear to be large enough to be in the prethermal regime, but the $U(1)$ symmetry-breaking is small. The rotation of the $^{19}$F nuclear spins is observable through its effect on the $0.13\%$ of the Ca atoms that are the $^{43}$Ca isotope.

A more recent experiment \cite{Autti18} similarly observes the rotation of the spins of $^3$He atoms in the $^3$He-B superfluid state. Again, the slow decay of the oscillations is due to weak breaking of the spin-rotational symmetry, and the field is not large enough to be in the exponentially-increasing regime, unlike in a prethermal continuous time crystal.

\section{Theory of Time-Translational Symmetry Breaking and Time Crystals}
\label{sec:TTSB}

\subsection{Spontaneous symmetry-breaking out of equilibrium}
\label{subsec:ssb_ooe}

In the previous section, we defined the context in which we wish to discuss phases of matter, namely many-body steady states in isolated quantum systems. In particular, if the many-body steady state does not respect the symmetries of the applied Hamiltonian, then we say that the symmetry is spontaneously broken.

If we now consider a periodically driven Hamiltonian with $H(t+T) = H(t)$, then the time-dependent Hamiltonian has a discrete time-translation symmetry. A discrete time crystal is a system where this discrete time-translation is spontaneously broken. More precisely,
\begin{quote}
A system in a many-body steady state for a periodically driven Hamiltonian with period $T$ is a \emph{discrete time crystal} if the expectation values of local observables are not $T$-periodic.
\end{quote}
For MBL discrete time crystals, the state state survives to infinite times; for prethermal discrete time crystals, to exponentially-long times. In the latter case, a time crystal will only occur for initial states with energy density below a critical value.

Another way to say this is that local observables have a longer period than the drive. In this review, we will be focusing on the case in which observables have a period $nT$ that is a multiple $n$ of the period of the drive.
Equivalently, the system has a ``subharmonic response'' to the drive since observables have fractional frequency $\Omega/n$.
In the simplest case, $n=2$, this is amounts to ``period doubling''. It is important, however, to distinguish the subharmonic response in time crystals from some phenomena that superficially might seem similar, as we discuss further in the next subsection.


\subsection{Non-trivial features of  time crystals}
As we stated previously, we have not attempted to give a restrictive definition of ``many-body state''; accordingly, we do not have a restrictive definition of time crystal. Nevertheless, we wish to emphasize certain features of time crystal phenomenology that we feel should be considered essential if the ``time crystal'' label is to be applied.

Firstly, as we have already stated, a time crystal is a phase of matter, not some finely tuned point in parameter space. Hence, the qualitative features of the steady state should be stable to perturbations of the Hamiltonian that respect time-translation symmetry, in particular to adding many-body interactions. A collection of uncoupled spins precessing in a magnetic field does not respect the time-translation symmetry, but as soon as the spins are coupled together, they will generically decohere and thermalize, and eventually the expectation values of local observables will approach values that are constant in time. Thus, uncoupled spins do \emph{not} constitute a time crystal. They are a highly unstable point, which can be perturbed into any one of many different time-translation-invariant phases with a suitable choice of interaction. By the same token, no extra symmetries (in addition to time-translation symmetry) are needed to stabilize a time crystal, unlike in pure dephasing models of spins.

Secondly, spontaneous symmetry breaking is intimately connected with a concept of ``rigidity''.  This means that the system should have many locally coupled degrees of freedom so that a notion of spatial dimension and thermodynamic limit can be defined \cite{Yao18b}, but in the spontaneous symmetry breaking phase all these degrees of freedom should get locked together into a symmetry-breaking order parameter configuration that has long-range order in both space and time.


Even within the constraints of the criteria mentioned above, it turns out that general classical dynamical systems can still exhibit rigid subharmonic responses.
The reason for this is that the dynamics about fixed points can be strongly damped so that perturbations to either the state or the dynamics decay rapidly; owing to the presence of such \emph{contractive} dynamics, many-body subharmonic responses have been observed in a multitude of systems including: Faraday wave instabilities \cite{cross1993pattern},  driven charge density wave materials  \cite{brown1984subharmonic, brown1985harmonic, tua1985dynamics, sherwin1985complete, balents1995temporal} and Josephson junction arrays \cite{lee1991subharmonic, yu1992fractional}. In this review, as we have already mentioned, we focus instead on obtaining time crystals in isolated quantum systems, which evolve unitarily without dissipation. In addition, dissipation caused by coupling to a reservoir should also come with noise caused by fluctuations in the reservoir. The stability of the subharmonic response of a damped system to such fluctuations is an open question \cite{Yao18a}.

\subsection{A prototypical model: the MBL discrete time crystal}
\label{sec:MBL}
Here we discuss a discrete time crystal that is stabilized by the presence of strong disorder leading to many-body localization (row 1, Figure 3).
To be specific, consider the following disordered spin model \cite{Khemani16,Else16b,Else17a,Yao17}.
We will define the Hamiltonian $H(t)$ by specifying it on the interval $[0,T)$ and imposing
periodicity $H(t+T) = H(t)$. The simplest stroboscopic Floquet Hamiltonian (with total evolution time $T=t_1 +t_2$)  takes the form (Fig.~4a):
\begin{equation}
H(t) = 
\begin{cases}
{H_1}\, , &\text{for } 0 \leq t < t_1\\
{H_2}\, , &\text{for } t_1 \leq t < t_1 + t_2
\end{cases}
\label{MBLham1}
\end{equation}
with time-independent Hamiltonians $H_1$, $H_2$ given by
\begin{align}
H_1 &= -\sum_{\langle i,j\rangle} J_{ij} \sigma^z_i \sigma^z_{j} -
\sum_i \left( h^z_i \sigma^z_i + h^y_i \sigma^y_i + h^x_i \sigma^x_i \right)\cr
H_2 &= \frac{\pi}{2 t_2}g \sum_i \sigma^x_i.
\label{MBLham2}
\end{align}
and $\vec{\sigma}$ being Pauli spin operators.
In order for this model to exhibit MBL in an appropriate regime, we choose the $J_{ij}, h_i^z, h_i^y, h_j^y$ from some random distributions. Moreover, it will turn out that this model exhibits a time-crystalline phase.

\begin{figure}[t]
    \includegraphics[width=0.9\textwidth]{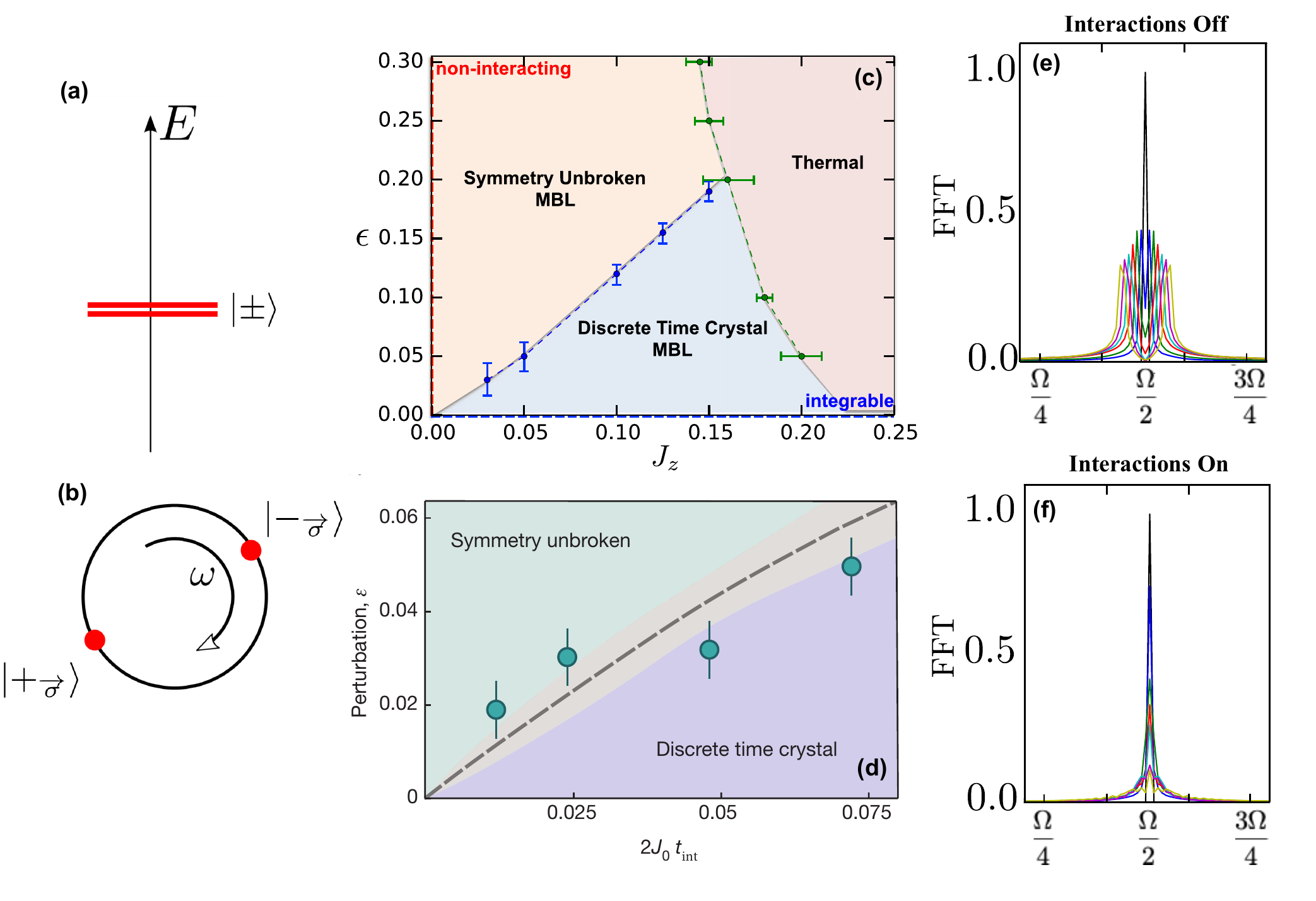}
    \caption{The eigenstate doublets associated with spontaneously breaking (a) an Ising symmetry, and (b) time-translation symmetry (with period-doubling). In the latter case, we have assumed that the system is periodically driven, so the quasi-energy $\omega$ is $\Omega$-periodic, where $\Omega = 2\pi/T$ is the angular frequency of the drive, and therefore is drawn on a circle. The two eigenstates in the multiplet are separated by $\Omega/2$ in quasienergy. (c) Schematic phase diagram associated with an MBL discrete time crystal. The underlying model (taken from \cite{Yao17}) is similar to Eq.~\eqref{MBLham2}, where $J_z$ is the mean value of the diordered  interaction strength and $\epsilon=g-1$ is the $\pi$-pulse imperfection. For large $\epsilon$ the MBL discrete time crystal melts into a symmetry unbroken phase \cite{Yao17,berdanier2018strong}, while for  large $J_z$, the disorder is not strong enough to localize the system, leading to a thermal phase. (d) This qualitative phase diagram is directly observed in small system trapped ion experiments described in Section \ref{sec:trappedions}. (e) The subharmonic response of the discrete time crystal is most easily observed in Fourier space. For the $n=2$ time crystal, in the absence of interactions, the subharmonic response lacks rigidity and is unstable to arbitrarily small perturbations $\epsilon$. (f) With interactions on, the many-body system synchronizes and exhibits a rigid subharmonic peak at $\omega/2$ despite the presence  of imperfections in the $\pi$-pulse.
    }
    \label{fig:dtc}
  \end{figure}

As we has emphasized, time crystals are a phenomenon that is stabilized by many-body interactions. To illustrate this, first consider the case where the spins are completely decoupled, $J_{ij} = 0$. Then we can just focus on the dynamics of a single spin $i$. If we also set $h_i^x = h_i^y = 0$, then the dynamics over a single time cycle is then given by
\begin{equation}
\label{uncoupled_UF}
U_{\text{F}} = e^{-i\pi g \sigma^x} e^{ih_i^z t_1 \sigma^z}.
\end{equation}
In the ideal case $g=1$, we find that $e^{ih_i^z t_1 \sigma^z}$ conserves the $z$ component of the magnetization, $\langle \sigma^z\rangle$, whereas $e^{-i\pi g \sigma^x}$ exactly flips it. Thus, if we start from a state initially polarized in the $z$ direction, successive applications of $U_{\text{F}}$ just flip the polarization direction each time, which looks like a subharmonic response at frequency $\Omega/2$. (In fact, one can show that $U_F$ is a rotation by angle $\pi$ about some tilted axis, so the same subharmonic response is exhibited for nearly any initial state). However, for $g = 1+\epsilon$, \eqnref{uncoupled_UF} now becomes a rotation by angle $\theta$ about some tilted axis, for $\theta$ not quite equal to $\pi$. In other words, the frequency of this subharmonic response varies continuously as a function of $g$, thus lacking the ``rigidity'' we normally associate with a spontaneously broken discrete symmetry. This lack of rigidity can be seen in Figure \ref{fig:dtc}, which depicts the Fourier transform of the response of the system. With $J_{ij} = 0$, the system exhibits a beat frequency as we tune away from $g=1$. This nonrigid subharmonic response is an artifact of the non-interacting limit.

By contrast, we expect that in an \emph{interacting} system, the only way a subharmonic response could occur is through the mechanism of many-body spontaneous symmetry breaking, and consequently it will be rigid. To illustrate this, consider the limit $J_i^x = h_i^y = 0$, but we choose some fixed $J_{ij}^z \neq 0$. Then the evolution over one time cycle becomes
\begin{equation}
\label{eq:coupled_UF}
U_{\text{F}} = e^{-i\pi g \sum_i \sigma^x} \exp\left(i\sum_i ih_i^z t_1 \sigma^z + i \sum_{\langle i,j\rangle} J_{ij} t_1 \sigma_i^z \sigma_j^z\right).
\end{equation}
If we tune $g$ exactly to 1, we again find that $e^{-i\pi g \sum_i \sigma^x_i} = \prod_i \sigma_i^x := X$ flips $\langle \sigma_i^z \rangle$ at each spin, whereas the rest of the evolution conserves it, so we see a subharmonic response at $\Omega/2$. But with interactions, this subharmonic response is \emph{stable}; it persists, even at infinite times, through a finite window of $g$ surrounding 1, and in fact is stable to any perturbation whatsoever in the driving Hamiltonian $H(t)$, provided that it remains $T$-periodic. Figure \ref{fig:dtc} shows how, unlike the non-interacting case, the sharp peak in the Fourier spectrum at $\omega = \Omega/2$ persists even as $g$ is tuned away from 1.

What causes the difference between the $J_{ij} = 0$ and $J_{ij} \neq 0$ cases? Here it is instructive to look at the spectrum (eigenstates and eigenvalues) of the Floquet evolution operator $U_F$. When the spins are uncoupled ($J_{ij} = 0$), $U_F$ is just a tensor product over the individual sites, so the eigenstates are all product states (or at least, can be chosen to be product states; in fact $U_{\text{F}}$ has a massive degeneracy when $J_{ij}=0$ and $g=1$, which is lifted for $g \neq 1$.) On the other hand, consider the interacting case in which $U_F$ has the form given in \eqnref{eq:coupled_UF} at $g=1$. For simplicity, we also set $h_i^z = 0$. Then the eigenstates of $H_{\mathrm{Ising}} = -i\sum_{\langle i,j\rangle} J_{ij} \sigma_i^z \sigma_j^z$ can be labelled by numbers $\sigma_i = \pm 1$, such that $\sigma_i^z \ket{\vect{\sigma}} = \sigma_i \ket{\vect{\sigma}}$. We find that $U_{\text{F}} = X e^{-iH_{\mathrm{Ising}} t_1}$ is block-diagonal in the eigenstates of $H_{\mathrm{Ising}}$, since it preserves the subspace spanned by a state $\ket{\vect{\sigma}}$ and its oppositely magnetized state $\ket{-\vect{\sigma}}$, and within this subspace it acts as
\begin{align}
U_{\text{F}} &= e^{-iE_{\vect{\sigma}} t_1}
 \left(\ket{\vect{\sigma}}\bra{-\vect{\sigma}} + \ket{-\vect{\sigma}} \bra{\vect{\sigma}}\right), \label{eq:sigma_subharmonic} \\
  &= e^{-i E_{\vect{\sigma}} t_1}( \ket{+}_{\vect{\sigma}}\bra{+}_{\vect{\sigma}} \, - \, \ket{-}_{\vect{\sigma}} \bra{-}_{\vect{\sigma}})
\end{align}
where $E_{\vect{\sigma}}$ is the energy of $\ket{\vect{\sigma}}$ (and also, by symmetry, $\ket{-\vect{\sigma}}$) under $H_{\mathrm{Ising}}$, and we have defined
$\ket{\pm}_{\vect{\sigma}}
\equiv \frac{1}{\sqrt{2}}( 
\ket{\vect{\sigma}} \pm \ket{-\vect{\sigma}} 
)$.
Thus, $\ket{\pm_{\vect{\sigma}}}$ are eigenstates of $U_{\text{F}}$. The important thing about these eigenstates is that they are \emph{cat states}; that is, the superposition of two macroscopically distinct states. This property of the eigenstates coming in cat state pairs, which are separated in quasi-energy by exactly $\Omega/2$, since their eigenvalues under $U_F$ are $\pm e^{-i E_{\vect{\sigma}} t_1}$ [see Figure \ref{fig:dtc}(b)] is the signature of a period-doubling time-crystal phase, as we will explain in more detail in the next subsection.
Moreover, this property is robust, in the sense that perturbations cannot alter this property provided that the localization length remains finite. Stability to perturbations usually requires a energy gap. However, MBL can similarly provide stability since perturbations can only cause local rearrangements of the eigenstates on the scale of the
localization length, which cannot alter the fact that eigenstates are cat states; for details, see Ref. \cite{Else16b}.
Thus, this property characterizes a stable phase of matter, the discrete time-crystal phase.

Intuitively, it is clear that the property of the eigenstates being cat states is related to the subharmonic response, because cat states are not observable in the laboratory, whereas the more physical $\ket{\vect{\sigma}}$ states discussed above exhibit subharmonic evolution [see \eqnref{eq:sigma_subharmonic}]. It was argued in Ref.~\cite{vonKeyserlingk16c} from the general phenomenology of MBL systems that, when time evolving a system in the time-crystal phase, for a generic physical initial state the expectation values of local observables will eventually oscillate with frequency $\Omega/2$ at late times. This corresponds to a many-body steady state as defined in Section \ref{sec:phases_ooe}, so this model really is a discrete time crystal as defined in Section \ref{subsec:ssb_ooe}.

Finally, let us briefly mention how the time-translation symmetry breaking interplays with the standard MBL phenomenology. Generally, a MBL system is characterized by the existence of a complete set of quasi-local integrals of motion; for a spin-half system, as we are considering here, this is a set of spin-half operators $\tau_i$, each of which is supported near the site $i$ (with exponential tails) and commutes with the Floquet evolution operator $U_F$ (in the case of Floquet systems) and the other $\tau_i$'s. ``Completeness'' means that simultaneous eigenstates of all the $\tau_i$'s are non-degenerate. It is these local conserved quantities that allow an MBL system to retain memory of its initial state forever instead of thermalizing.

When a symmetry is spontaneously broken in an MBL system, this corresponds to the $\tau_i$'s not commuting with the symmetry; this allows the constant in time expectation values of these operators to serve as an ``order parameter'' for the spontaneous symmetry breaking. We can see this for the MBL DTC by observing that if $U_F$ is the Floquet evolution operator \eqnref{eq:coupled_UF}, with $g=1$, then $(U_F)^2$ commutes with $\sigma_i^z$ for each site $i$. Hence for this exactly solvable point, we can take $\tau_i = \sigma_i^z$, and in the presence of perturbations the $\tau_i$'s will be some dressed version of this. Crucially, though $\tau_i$ commutes with $(U_F)^2$, it does not commute with $U_F$; in fact, we have $U_F \tau_i U_F^{\dagger} = -\tau_i$. Since $U_F$ is the generator of time-translation symmetry, this reflects the spontaneously broken time-translation symmetry.


  
\subsection{Eigenstate definitions of spontaneous symmetry breaking and time crystals}
\label{sec:eigenstates}
As we saw in the previous section, the eigenstate properties of the MBL discrete time crystal are very striking. In fact, it is reasonable to \emph{define} time crystals in isolated quantum systems in terms of these properties. The advantage of such an approach is that it gives a sharper definition than the somewhat vague notion of a ``many-body steady state'' which we introduced in Section \ref{sec:phases_ooe}, and also allows closer connections to be drawn with the notion of spontaneous symmetry breaking in equilibrium (thus building on the observation in Section \ref{sec:locprotected} that eigenstates lead to an explicit connection between MBL systems and zero-temperature equilibrium phases of matter). The disadvantage, of course, is that eigenstates are usually not experimentally accessible states, so such a definition will necessarily be more theoretical. In this section, we outline such an eigenstate approach to the definition. We leave it as an open question whether such a definition is always equivalent to the one in terms of many-body steady states from Section \ref{subsec:ssb_ooe}.

\subsubsection{Spontaneous symmetry breaking generally}
\label{sec:defn_general}
As motivation, let us first recall how spontaneous symmetry breaking works at zero temperature; that is, in the ground state of a static Hamiltonian $H$. The classic example is the ground state subspace of an Ising ferromagnet (which has a $\mathbb{Z}_2$ Ising spin-flip symmetry) is degenerate and spanned by a pair of spin-polarized states in which the spins have a net magnetization in the up direction or the down direction (in the limit of vanishing transverse field, they are fully-polarized in the up or down direction); we call these states $\ket{\uparrow}$ and $\ket{\downarrow}$, and they are related by the Ising symmetry. On any finite system, however, there is some tunneling amplitude between these two states, as a consequence of which the true eigenstates are the symmetric and anti-symmetric combinations $\ket{\pm} = \frac{1}{\sqrt{2}}(\ket{\uparrow} \pm \ket{\downarrow})$, which are nearly degenerate with an energy separation that is exponentially small in the system size. (Since the Hamiltonian commutes with the symmetry, if the eigenstates are non-degenerate they must be invariant under the symmetry, as the $\ket{\pm}$ states are). The signature of the spontaneous symmetry breaking is that, while the symmetry-breaking states $\ket{\uparrow}$ and $\ket{\downarrow}$ are short-range correlated states, the invariant combinations $\ket{\pm}$ are long-range correlated ``cat states''; for example, the connected correlator $\langle \hat{m}(x) \hat{m}(y) \rangle - \langle \hat{m}(x) \rangle \langle \hat{m}(y) \rangle$ remains nonzero even when $|x - y| \to \infty$, where $\hat{m}(x)$ is the local magnetization operator.

Although this eigenstate multiplet structure is most familiar in ground states, the same structure is found in highly excited states for systems that exhibit spontaneous symmetry breaking at finite energy density; this is true both in systems that obey ETH \cite{Fratus_1505,Mondaini_1512,Fratus_1611} and in systems with MBL \cite{Huse_1304}. This gives us a definition of spontaneous symmetry breaking out of equilibrium in isolated quantum systems:

\begin{quote}
Let $H$ be the time-independent Hamiltonian of an isolated quantum system, or else let $U_{\text{F}}$ be the Floquet evolution operator corresponding to a time-periodic Hamiltonian. Suppose $H$ or $U_{\text{F}}$ has a symmetry operation represented by a unitary or anti-unitary operator $u$. Then the symmetry is spontaneously broken in an eigenstate $\ket{\psi}$ of $H$ (or $U_{\text{F}}$) if there is no linear combination of finitely many eigenstates of $H$ (or $U_{\text{F}}$), each of which has approximately the same energy (or quasi-energy) as $\ket{\psi}$, such that the linear combination is both a short-range correlated state and invariant under $u$ (up to a global phase). Here by ``approximately the same energy'', we mean that the energy difference is exponentially small in the system size.
\end{quote}

There is another way to formulate the definition of spontaneous symmetry breaking in terms of ``off-diagonal long-range order''. Let $\hat{o}(x)$ be a family of local operators (usually called the ``order paramater'') supported at different positions in space, such that $u \hat{o}(x) u^{-1} = e^{i \alpha} \hat{o}(x)$ for some phase factor $e^{i\alpha} \neq 1$. Then we say that $u$ is spontaneously broken in a $u$-invariant (up to global phase) eigenstate $\ket{\psi}$ if
\begin{equation}
\bra{\psi} \hat{o}(x) \hat{o}(y) \ket{\psi} \nrightarrow 0 \quad \mbox{as $|x - y| \to \infty$}.
\end{equation}
This can be shown to be equivalent to the definition above if we supplement the latter by a few extra technical assumptions; we give the details in Appendix \ref{appendix:multiplets}. This formulation of the definition is particularly convenient in numerics, where the formulation in terms of linear combinations of eigenstates can be difficult to check since the (quasi-)energy spacing of eigenstates is generally exponentially small in the system size.

\subsubsection{Discrete time crystals}
\label{sec:defn_dtc}
 As we already saw in the model of an MBL discrete time crystal discussed in \eqnref{MBLham2}, it also exhibits a similar multiplet structure to the one described just above. The main difference is that whereas in an Ising symmetry breaking phase, for example, the paired eigenstates $\ket{\pm}$ are nearly degenerate, in the discrete time crystal they are separated by a quasi-energy very close to (exactly equal to in the thermodynamic limit) $\Omega/2$, where $\Omega$ is the driving frequency. More generally, the quasi-energy separation of eigenstate multiplets is related to the fractional frequency response; for example, for a time crystal that responds with period $NT$, there would be $N$ eigenstates separated by quasi-energy $\Omega/N$.

In any case, the definition above of spontaneous symmetry breaking applies equally well to discrete time translation symmetry; we just have to remember that the generator of the symmetry is $U_{\text{F}}$ itself. Hence, we obtain the definition
\begin{quote}
Let $U_{\text{F}}$ be the Floquet evolution operator corresponding to a time-periodic Hamiltonian. Then the discrete time-translation symmetry is spontaneously broken in an eigenstate $\ket{\psi}$ of $U_{\text{F}}$ if there is no linear combination of finitely many eigenstates of $U_{\text{F}}$, each of which has approximately the same quasi-energy as $\ket{\psi}$, such that the linear combination is a short-range correlated state.
\end{quote}

We can also try to formulate a definition in terms of off-diagonal long-range order. Recall that for this we wanted to consider operators that transform under the symmetry as $u \hat{o}(x) u^{-1} = e^{i\alpha} \hat{o}(x)$. However, here $u = U_F$, and since Heisenberg time evolution generally causes local operators to spread, it is not clear that there will generally be local operators satisfying this condition\footnote{Ref.~\cite{Khemani_1612} constructed quasi-local operators satisfying this condition for the MBL discrete time crystal discussed above, but they are unlikely to exist for the prethermal time crystals discussed below because operators spread much more rapidly in the absence of MBL.}.
 Instead, in Appendix \ref{appendix:multiplets} we show (given some extra technical conditions) that a sufficient condition for discrete time-translation symmetry to be spontaneously broken in an eigenstate $\ket{\psi}$ is that there exists a family of local operators $\hat{o}(x)$ such that the unequal time correlator at large separations fails to be $T$-periodic in the time difference, or more precisely
 \begin{equation}
 C(T; x,y) - C(0; x,y) \nrightarrow 0 \quad \mbox{as $|x - y| \to \infty$},
 \end{equation}
 where \begin{equation}
 C(nT; x,y) = \bra{\psi} U_{\text{F}}^n \hat{o}(x) U_{\text{F}}^{-n} \hat{o}(y) \ket{\psi}.
 \end{equation}
 This is a discrete version of the diagnostic proposed in Ref.~\cite{Watanabe15}, and was also referred to as ``spatiotemporal long-range order'' in Ref.~\cite{vonKeyserlingk16c}.

\subsection{Prethermal discrete time crystal}
 \begin{figure}[t]
    \includegraphics[width=5in]{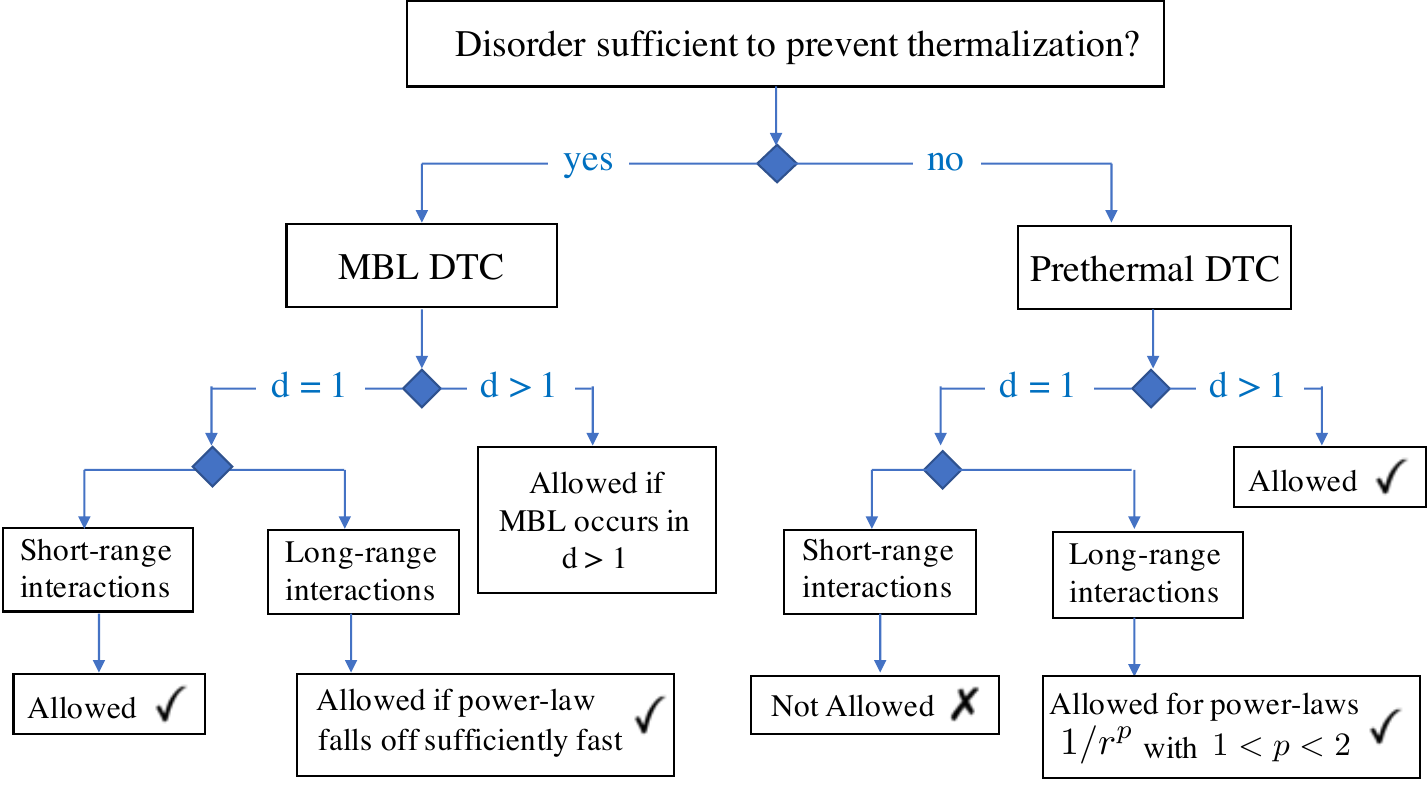}
    \caption{Flowchart clarifying the key differences between the MBL and prethermal discrete time crystals (DTC). Especially striking is the following dichotomy: The MBL DTC is well established in $d=1$ for systems with short range interactions, while long-range interactions tend to destabilize localization. On the other hand, the prethermal DTC is forbidden in $d=1$ with short range interactions, but can be stabilized in the presence of long-range interactions. At its core, this fact owes to the absence of finite temperature Ising symmetry breaking in $d=1$ with short range interactions.
    Moreover, this difference again highlights the fact that MBL eigenstates behave as ``thermal states'' with zero temperature, while the prethermal eigenstates of $H_F$ behave as finite temperature thermal states.
    }
    \label{Fig:flowchart}
  \end{figure}

\label{sec:prethermal_dtc}

As we have already mentioned, disorder is not the only way to stabilize phases of matter in driven systems; we can also consider systems exhibiting a Floquet prethermal regime as described in Section \ref{subsec:crypto}. Here we will describe specifically how to get a discrete time crystal in this regime (row 2, Figure 3). Recall that in the Floquet prethermal regime described in Section \ref{subsec:crypto}, there exists a quasi-local unitary time-independent change of frame $\mathcal{U}$ such that the Floquet evolution can be approximated according to
\begin{equation}
\widetilde{U}_F := \mathcal{U} U_F \mathcal{U}^{\dagger} \approx \widetilde{U}_F^{\mathrm{approx}} = X e^{-iD T} \label{eq:UFapprox}
\end{equation}
where $X$ is a unitary satisfying $X^N = 1$ for some positive integer $N$, and $D$ is a quasi-local Hamiltonian with $[D, X] = 0$. If we only observe the system at times $t = nNT$, it appears to be evolving under the time-independent Hamiltonian $D$, and we expect the expectation values of local observables to converge to those of a thermal state $\rho$ of $D$. The temperature of this thermal state is set by the expectation value of $D$ in the initial state, since $D$ is approximately conserved in the prethermal regime.

What one can imagine happening is that the $\mathbb{Z}_N$ symmetry generated by $X$ could be spontaneously broken in the state $\rho$. This turns out to give a time crystal if we allow ourselves to observe the system at times other than $t = nNT$. Indeed, letting $N=2$ for simplicity, spontaneously broken $\mathbb{Z}_2$ symmetry means there is an order parameter, represented by a local observable $\hat{o}$ with $X \hat{o} X^{\dagger} = -\hat{o}$, such that $\mathrm{Tr}(\hat{o} \rho) \neq 0$. If we now ourselves to observe the system at odd times, $t = (2n+1)T$, we find that
\begin{equation}
\left\langle \hat{o}\Bigl([2n+1]T\Bigr) \right\rangle \approx \mathrm{Tr}[\widetilde{U}_F^{\mathrm{approx}} \rho (U_F^{\mathrm{approx}})^{\dagger} \hat{o}] = \mathrm{Tr}(X^{\dagger} \hat{o} X \rho) = -\mathrm{Tr}(\hat{o} \rho) = -\bigl\langle \hat{o}(2nT) \bigr \rangle,
\end{equation}
which describes time-crystalline behavior. (Here we used the fact that $[D, \rho] = 0$ since $\rho$ describes a thermal state of $D$).

For a concrete model, we can use the same model in \eqnref{MBLham1} and (\ref{MBLham2}) that gave rise to an MBL DTC, where we suppress MBL by weakening the disorder; for example, we can remove the disorder completely so that all the couplings become translationally invariant. If we demand that $\Omega_0 := \mathrm{max}\{ |g-1|/t_2, |J_{ij}|, |J^x_{ij}|, |h^x_i|, |h^y_i|, |h^z_i| \} \ll \Omega = 2\pi/T$, then the conditions of the theorem discussed in Section \ref{subsec:crypto} are satisfied with $N=2$, $X = \prod_i \sigma_i^x$, and we find that $\widetilde{U}_{\text{F}} = \widetilde{U}_{\text{F}}^{\mathrm{approx}} + O(e^{-\Omega/\Omega_0})$. We can also compute $D$ to leading order in $\Omega_0/\Omega$, which turns out to be
\begin{equation}
\label{eq:isingD}
D = -\sum_{\langle i,j \rangle} (J_{ij} \sigma_i^z \sigma_j^z + J_{ij}^x \sigma_i^x \sigma_j^x) +
 \sum_i \left[\frac{\pi(g-1)}{2T} - h_i^x\right] \sigma_i^x + O[(\Omega_0/\Omega)]^2.
\end{equation}
Notice that this commutes with $X$ as expected (the Ising symmetry-breaking terms in the original Hamiltonian have been rotated away by the change of frame $\mathcal{U}$). If $J_{ij}$ is large compared with rest of the couplings in \eqnref{eq:isingD}, then we expect $D$ to have an Ising spontaneous symmetry-breaking phase at low temperatures in spatial dimension $d > 2$, which will rise to time-crystalline behavior as previously described. However, at very late times $t_* = O(e^{\Omega/\Omega_0})$, the system will start to absorb energy from the drive, and the prethermal description will cease to be valid.

It should be obvious from the above discussion that the prethermal time crystals satisfy the definition of time crystal from Section \ref{subsec:ssb_ooe}, because the state $\rho$ is a many-body steady state (at least until time $t_*$). Let us show that it also satisfies the eigenstate definition from Section \ref{sec:eigenstates}. First of all, to describe the spontaneous symmetry-breaking in the prethermal regime we should consider the eigenstates of the approximate Floquet evolution operator $\widetilde{U}_{\text{F}}^{\mathrm{approx}} = Xe^{-iDT}$. (Or rather, its unrotated version $U_F^{\mathrm{approx}} = \mathcal{U}^{\dagger} \widetilde{U}_F^{\mathrm{approx}} \mathcal{U}$, but one can check that conjugation by local unitaries does not affect the time crystal definition). Moreover, if $D$ spontaneously breaks the symmetry generated by $X$ at finite temperature, then the corresponding finite-energy eigenstates of $D$ come in pairs $\ket{\uparrow}$ and $\ket{\downarrow}$ with opposite magnetization \cite{Fratus_1505,Mondaini_1512,Fratus_1611}. Hence we conclude that the corresponding eigenstates of $\widetilde{U}_F^{\mathrm{approx}}$ are $\frac{1}{\sqrt{2}}(\ket{\uparrow} \pm \ket{\downarrow})$, separated in quasi-energy by $\Omega/2$, which indeed satisfy the definition of a discrete time crystal from Section \ref{sec:eigenstates}.

We conclude by summarizing the main differences betweeen the prethermal DTC and the MBL DTC (see also the flowchart Figure \ref{Fig:flowchart}). The prethermal DTC persists only until the exponentially long heating time, whereas MBL persists forever (in a completely isolated system). The prethremal DTC has a dependence on the initial state; its energy density with respect to $H_F = \mathcal{U}^{\dagger} D \mathcal{U}$ must be sufficiently low that the thermal state with respect to $D$ at that energy density spontaneously breaks the symmetry. By contrast, the MBL DTC will exist for any initial state (provided that it is short-range correlated). Finally, the MBL DTC can exist in one dimension (and possibly in higher dimensions as long as MBL itself can exist in higher dimensions), but with short-range interactions, the prethermal DTC requires spatial dimension $d > 2$, because the Mermin-Wagner theorem forbids a finite-temperature symmetry-breaking phase transition for a discrete symmetry. With long-range interactions, that scale with distance as $\sim 1/r^\alpha$ with $1 < \alpha < 2$, one can potentially have a finite-temperature phase transition in one spatial dimension. On the other hand, the prethermalization theorems described in Section \ref{subsec:crypto} do not apply to such long-ranged interactions. Nevertheless, there is numerical evidence that a prethermal regime still exists in this case and hosts a prethermal DTC \cite{Machado_1708, Machado_InPreparation}.

\section{Other Long-Lived Non-Equilibrium Regimes with Time-Crystalline Signatures}
\label{sec:models+expts}

  \begin{figure}[t]
    \includegraphics[width=0.9\textwidth]{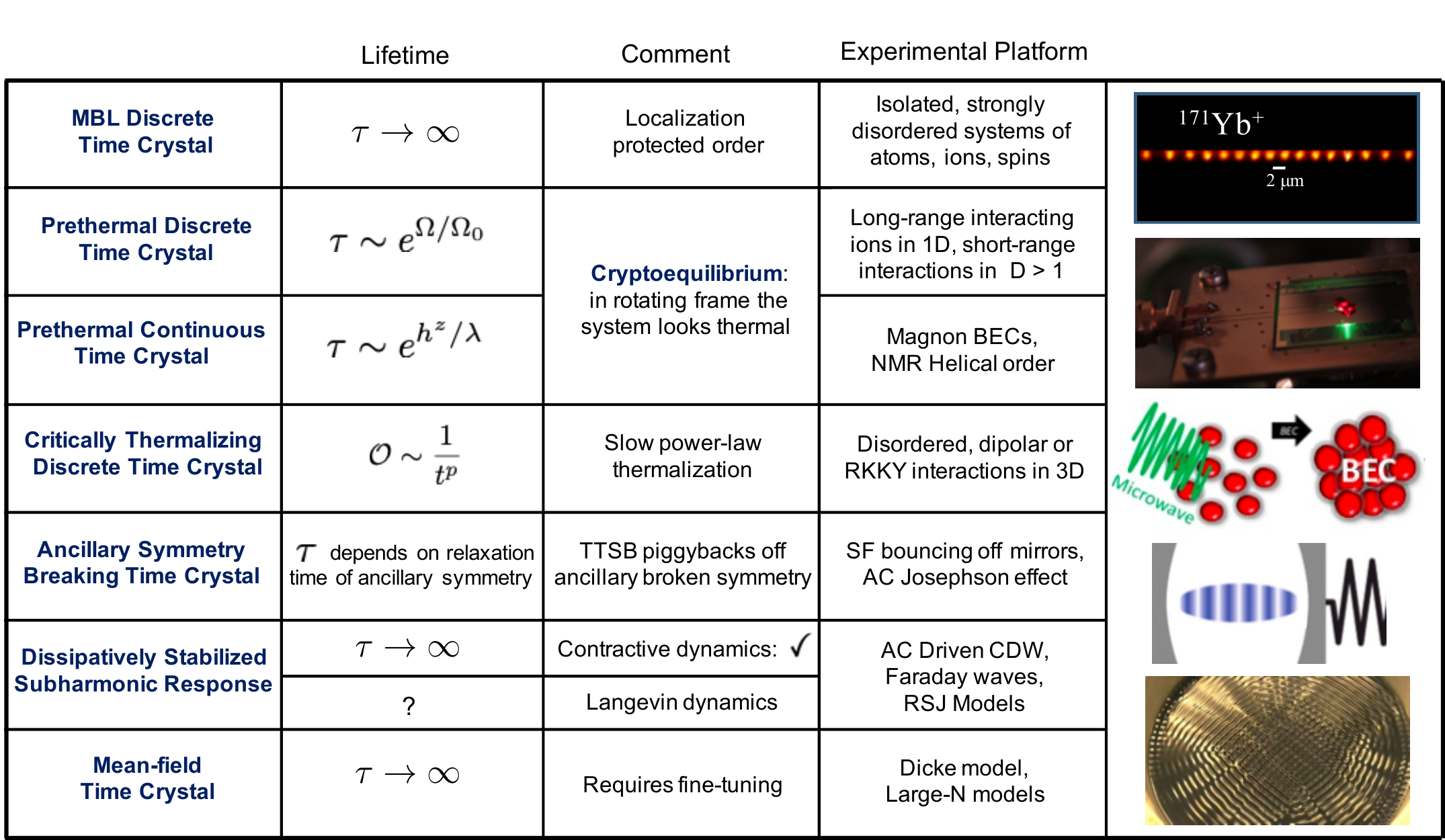}
    \caption{Zoology of time crystals. The MBL DTC represents an example of localization protected order and can occur in isolated AMO platforms with strong disorder. The prethermal time crystals (both discrete and continuous) have exponentially long lifetimes controlled by an external field. In the case of Magnon Bose-Einstein condensates and helical order in NMR systems, although we have placed them in the prethermal category, if the field strength is small (as discussed in Section \ref{sec:prethermal-continuous}), they are more appropriately labeled as ancillary time crystals. Both MBL and prethermal time crystals have the property that their eigenstates look like ground states or finite-temperature states of stationary systems.  For the case of critical time crystals, simple resonance counting arguments suggest that the DTC order $\mathcal{O}$ could decay slowly as a power-law, although experiments have not yet clearly observed this behavior. Ancillary time crystals are so-named because their time translation  symmetry breaking (TTSB) depends on the existence of another ancillary symmetry being broken (i.e.~$U(1)$ symmetry breaking in a superfluid). For dissipative open systems, we distinguish two cases. The first is that of contractive dynamics where subharmonic responses can be stabilized by simply damping away all perturbations. The second and more subtle case is that of Langevin dynamics where dissipation comes hand-in-hand with noise \cite{Yao18a}. Included in this case is the situation described in Section \ref{subsec:dissipative}, where a system is coupled to a cold bath. Finally,  we refer to mean-field time crystals as those models where either an all-to-all coupling or a large-$N$ limit enables a simplified few-body description  of the dynamics. It is unclear whether such an approach can survive heating effects even for an arbitrarily small integrability breaking perturbation  \cite{zhu2019dicke}. 
    }
    \label{Fig:Zoo}
  \end{figure}
  
In addition to the MBL and prethermal time crystals described in Section \ref{sec:TTSB}, there are a number of other strategies to impede thermalization.
In the context of these alternate strategies, there has recently been an explosion of both theoretical proposals and experiments in the broad landscape surrounding time crystals \cite{yu2018discrete,OSullivan18,nalitov2018optically,xu2018nutation,smits2018observation, cole2018kerr, surace2018floquet, ohberg2018quantum, liao2019dynamics,oberreiter2019subharmonic, efetov2019thermodynamic,dai2019truncated, cai2019imaginary}. 
In this section, we will focus on summarizing this zoology (Fig. \ref{Fig:Zoo}) with an eye toward clarifying the distinction between these systems and MBL/prethermal time crystals. We note that the phenomena described here will not necessarily satisfy the strict definition of time crystals discussed above. Therefore, one might want to describe them as ``time-crystalline signatures'', rather than true ``time crystals'' as such.



\subsection{Ancillary time crystals}

The discussion of prethermal continuous time crystals in Sec. \ref{sec:prethermal-continuous} is essentially the logic that was discussed
in Refs.~\onlinecite{Wilczek12,Volovik_1309,Watanabe15},
where it was pointed out that a superfluid
at non-zero chemical potential is a time crystal as a result of the well-known
time-dependence of the order parameter \cite{Pethick__08}.
However, there is an important difference: the U(1) symmetry is not a symmetry of
the Hamiltonian of the problem and, therefore, does not require
fine-tuning but, instead, emerges in the $h^z \rightarrow \infty$ limit, thereby
evading the criticism \cite{Nicolis2012,Castillo2014,Thies2014,Volovik_1309,Watanabe15}
that the phase winds in the ground state only if the U(1) symmetry is exact.

However, there are systems in which time crystal behavior \cite{sacha2015modeling,mizuta2018spatial,matus2018fractional} actually does ``piggyback'' off another
broken symmetry (row 5, Figure 3). This does require fine-tuning, since it is necessary to ensure that the system
posseses the ``primary'' symmetry, but such tuning may be physically natural
(e.g. helium atoms have a very long lifetime, leading to a U(1) symmetry).
The broken symmetry allows a many-body system to effectively become a few-body
system. Thus, time crystal behavior can occur in such systems for the same reason that oscillations
can persist in few-body systems. Oscillating Bose condensates (e.g. the AC Josephson effect and the
model of Ref.~\onlinecite{sacha2015modeling}) can, thus, be viewed as fine-tuned time crystals.
They are not stable to arbitrary time-translation symmetry-respecting perturbations; a perturbation
that breaks the ``primary'' symmetry will cause the oscillations to decay.
Indeed, most few-body systems are actually many-body systems in which a spontaneously-broken
symmetry approximately decouples a few degrees of freedom. A pendulum is a system of $~10^{23}$ atoms
that can be treated as a single rigid body due to spontaneously-broken spatial translational symmetry:
its oscillations owe their persistence to this broken symmetry, which decouples the center-of-mass
position from the other degrees of freedom. The non-trivial feature of the MBL and prethermal time crystals discussed above is that there is no microscopic symmetry other than time-translation symmetry which is spontaneously broken; the rigidity thus comes solely from the time-translation symmetry breaking.

\subsection{Mean-field time crystals}

Symmetry breaking is not the only context where a few-body description of a many particle interacting system can occur; a different context arises when mean-field is the correct treatment.
In this case, although the microscopic model can be characterized by many degrees of freedom, the behavior of the system can be captured via only a few degrees of freedom (row 7, Figure 3) \cite{chandran2016interaction,barnes2019stabilization, barfknecht2018realizing, PalS18,gong2018discrete}.



A classic example of this physics is captured by the Dicke model of $N$ two-level atoms interacting with a cavity photon mode \cite{gong2018discrete}.
By ``integrating'' out the photon degrees of freedom, the atomic system reduces to an all-to-all coupled spin model amenable to a mean-field treatment.
At the same time, because of the permutation symmetry of the atoms, the $N$ spins can be recast as a single large spin $S=N/2$ interacting with the photon field.
In this case, the observation of the time-crystalline behavior owes to the existence of a few-body description for the system rather than the stabilization of DTC order via interactions (as in the MBL and prethermal cases).

More specifically, from this few-body description one immediately observes the existence of two distinct phases controlled by the strength of the coupling between the atoms and the photon mode.
When the coupling is small, the ground state is given by the vacuum mode of the cavity and all the atoms in their ground state.
However, when the coupling is large, there are two degenerate ground states characterized by a coherent superposition of states of the atoms and a non-zero number of photons in the cavity---this corresponds to the celebrated superradiant state.
To this end, time crystalline behavior can be observed by engineering a protocol that rotates the system between the two degenerate ground states within each period.
However, as discussed above, when considering perturbations around this idealized protocol, the drive injects energy into the system and the superradiant regimes lose its stability at late time.
Nevertheless, by using a leaky cavity (i.e.~dissipation), where the photons can escape, the excess energy can be extracted from the system and it may be possible to stabilize the Dicke time crystalline behavior.

However, this simple picture survives only when the mean-field description is valid.
In the presence of generic interactions that break the mean-field description, one expects the time crystalline behavior to become less robust.
In this case, such generic interactions are not the crucial ingredient for stabilizing the time crystalline phase (as they are in the MBL and prethermal case) but rather serve as disrupting force moving the system away from the solvable Dicke limit \cite{zhu2019dicke}.
To this end, demonstrating the generic stability of such systems is an open question that may lead to new avenues for the realization of many body time crystalline order.

\subsection{Dissipative Stabilization, Open Systems and Classical Time Crystals}
\label{subsec:dissipative}

Thus far, we have defined a time crystal to be a phase of matter of a closed system. But suppose we relax this condition and consider open systems. Might we consider some open systems to be time crystals?
As we already saw in the Dicke model,
a dissipative bath (row 6, Figure 3)
can play a significant role in preventing the drive-induced heating of a Floquet system and thus stabilize a wider range of strongly interacting time crystalline phases \cite{Else17a,tucker2018shattered,OSullivan18,gambetta2019discrete,lledo2019driven,droenner2019stabilizing}.
In systems near thermal equilibrium, a bath that is itself in equilibrium will help the system to reach equilibrium.
But in a driven system, the bath can, instead, enable the system to reach a non-equilibrium steady-state.
Imagine coupling a bath very weakly to
a system that would, in isolation, be prethermal.
The bath could, conceivably, have no effect other than to counteract the slow heating of the system, thereby enabling the time crystal to survive to infinite times. In such a case, we would probably consider such an open system to be a time crystal. On the other hand, if the bath were to play the dominant role in stabilizing time crystal order, such as in the  extreme limit of purely contractive dynamics in which there
is damping but no noise, we would probably not consider this to be a time crystal \cite{Yao18a}. Indeed, open systems can have quite different physics than closed ones: for instance, even a zero-dimensional open system can undergo a quantum phase transition (e.g. in the Caldeira-Leggett model \cite{Leggett87}), and the entropy of an open system can decrease since entropy can be dumped into the bath.
It is an important open problem to determine if there is a sharp definition of a time crystal in open systems that retains the key features of the closed system time crystal. 
One possible hint along these lines relates to the physics of probabilistic (noisy) 1D cellular automata \cite{gacs2001reliable}. Since such automata are proven to be able to simulate a deterministic Turing machine, they can directly realize a period-doubled time crystal by implementing the ``program'' which flips all bits every cycle. If such a probabilistic cellular automata could then be \emph{faithfully} simulated by a classical Hamiltonian of coupled oscillators interacting with a Langevin bath, this would naturally qualify as a classical dissipative time crystal. 
Recent efforts have explored this possibility in a simple, generic model, but rather than finding a true classical time crystal, one observes thermally activated behavior, with an intriguing non-equilibrium phase transition \cite{Yao18a}.

\subsection{Critical Time Crystals}
\label{subsec:critical}

Finally, we focus on disordered long-range interacting systems where the power-law exponent matches the dimensionality of the system, leading to the phenomenon of \emph{critical thermalization} \cite{Ho17, Kucsko18}; importantly, this behavior is present in some of the platforms where time crystalline order was first observed \cite{ChoiS17, choi2019probing}.

The nature of this critical thermalization behavior can be illustrated via a simple resonance counting argument (which holds in the weak interaction limit with strong disorder in the local effective magnetic field).
Two spins can interact if the interaction between them is greater than difference between their local effective magnetic fields; when this occurs, we will say that they are in resonance.
When the power-law with which the interaction decays matches the dimension of the system, a degree of freedom at the origin will have probability $\propto 1/r$ to find a resonance in a thin spherical shell at radius $r$:
\begin{equation}
dP_\text{res} = \left(\frac{Jr^{-d}}{W} \right) \rho \times C_d r^{d-1} dr =  \frac{C_d J\rho}{W}\, \frac{dr}{r}
\end{equation}
where $J$ and $W$ are the interaction and disorder strengths, $d$ is the dimension, $C_d$ is a constant, and $\rho$ is the density of spins in the system. The probability that a spin at the origin interacts with a spin in a thin spherical shell at radius $r$ is then:
\begin{equation}
dP_\text{int} = (1-P_\text{int}) \frac{C_d J\rho}{W}\, \frac{dr}{r}
\end{equation}
The factor of $(1-P_\text{int})$ on the right-hand-side is the probability that the spin at the origin didn't already interact with another spin closer than $r$. Integrating over $r$, we find the probability that the spin at the origin hasn't interacted with any spin within radius $r$ is:
\begin{equation}
1-P_\text{int}(r) =  \left(\frac{a_0}{r}\right)^\frac{C_d J\rho}{W}
\end{equation}
At the same time, a resonance only becomes meaningful for the dynamics of the system at times later than the inverse of the interaction strength. At time $t$, the dynamics can only affect resonances that have separation less than $R^* = (tJ)^{1/d}$.
The deviation from equilibrium, at some time $t$, is captured by the probability that a spin at the origin hasn't interacted with a spin within radius $R^*$:
\begin{equation}
1-P_{\text{int}}({}R^*) =  \left(\frac{a_0}{tJ}\right)^\frac{J\rho}{dW}
\end{equation}   
In other words, the deviation from equilibrium decays as a power-law in time.

This behavior is quite distinct from the different thermalizing behaviors we considered so far. 
On the one hand, the system indeed approaches thermal equilibrium and thus does not have an infinitely long-lived non-ergodic phase such as MBL.
In fact, the long-range nature of the interactions present in such systems preclude them from exhibiting an MBL phase \cite{yao2014many}.
On the other hand, instead of a natural time scale related to the strength of the interactions, the system exhibits a scale free power-law approach to the equilibrium state.
The result is a long-lived regime during the approach to the featureless infinite temperature state, during which transient discrete time crystalline order can be observed (row 4, Figure 3).

At first sight, this behavior is reminiscent of prethermal time crystalline order, however we emphasize a few crucial distinctions.
Firstly, in prethermal time crystals, the DTC order depends crucially on the energy density of the initial state, which sets the temperature of the cryptoequilibrium thermal state of the prethermal regime.
In contrast, in critical time crystals, one can observe transient DTC order across the entire spectrum of the system.
Secondly, prethermal time crystal exhibit an \emph{exponentially} long thermalization time scale controlled by the frequency of the drive, but critical time crystals do not have such a simple dependence on the frequency and no parameter by which the heating time can be straightforwardly extended.

Finally, let us add a caveat. It is not clear whether the simple resonance counting arguments presented here survive the addition of many body interactions; more precisely, the theoretical analysis which leads to a power-law decay considers a pair depolarization process in which a pair of spins depolarize together via a resonance, but does not consider the effect of multi-spin (i.e.~beyond two-body) resonances. 

\section{Experimental Observations of Time Crystals}

In this section, our goal is to summarize several recent experiments \cite{ZhangJ17,ChoiS17,Rovny18a}, which have demonstrated a sufficient level of quantum control in order to observe certain features of discrete time-crystalline order. 
These experiments help to highlight the viewpoint that the subharmonic oscillations intrinsic to a time crystal can be observed in a number of different experimental platforms, each of which begs a different theoretical explanation. 
At the same time, limited by a combination of experimental noise and decoherence, the combined results of experiments to date point to the need for additional studies to truly demonstrate the presence of long-range order in both space and time. 
programmable
In each of the experimental platforms, there are several unique tools to control time crystalline behavior and to delay the onset of thermalization, including  disorder and long-range interactions. 
However, there are limitations as well, and these point to the following juxtaposition: systems with many degrees of freedom are plagued with inhomogeneities and limited individual control, while  systems with more control are necessarily smaller and have pronounced finite size effects.

\begin{figure}[t]
    \includegraphics[scale=0.5]{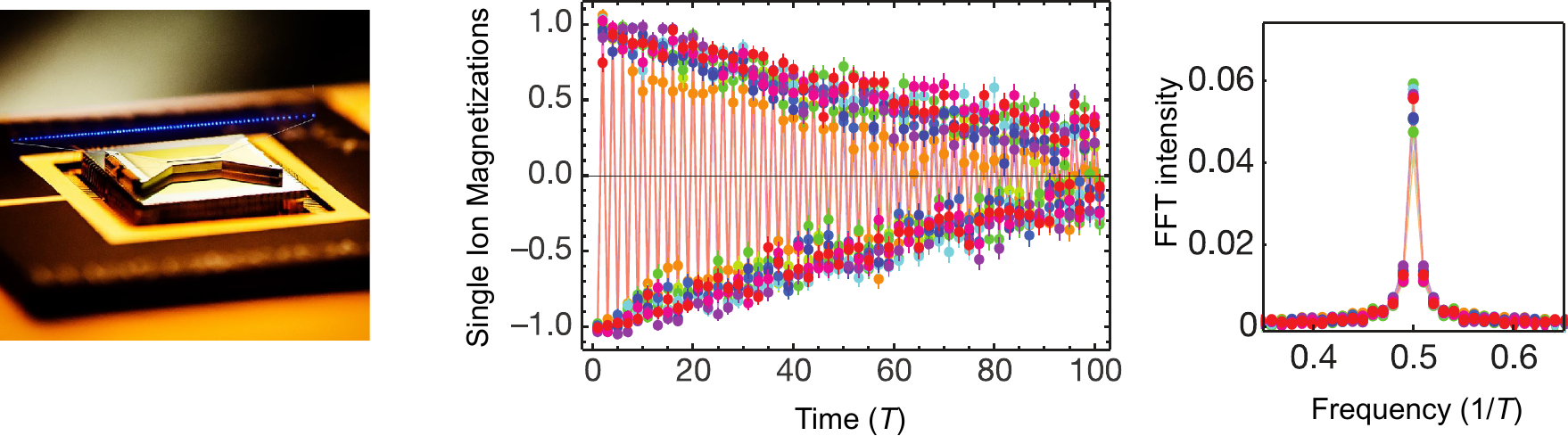}
    
    \vspace{0.5cm}
    
    \includegraphics[scale=0.5]{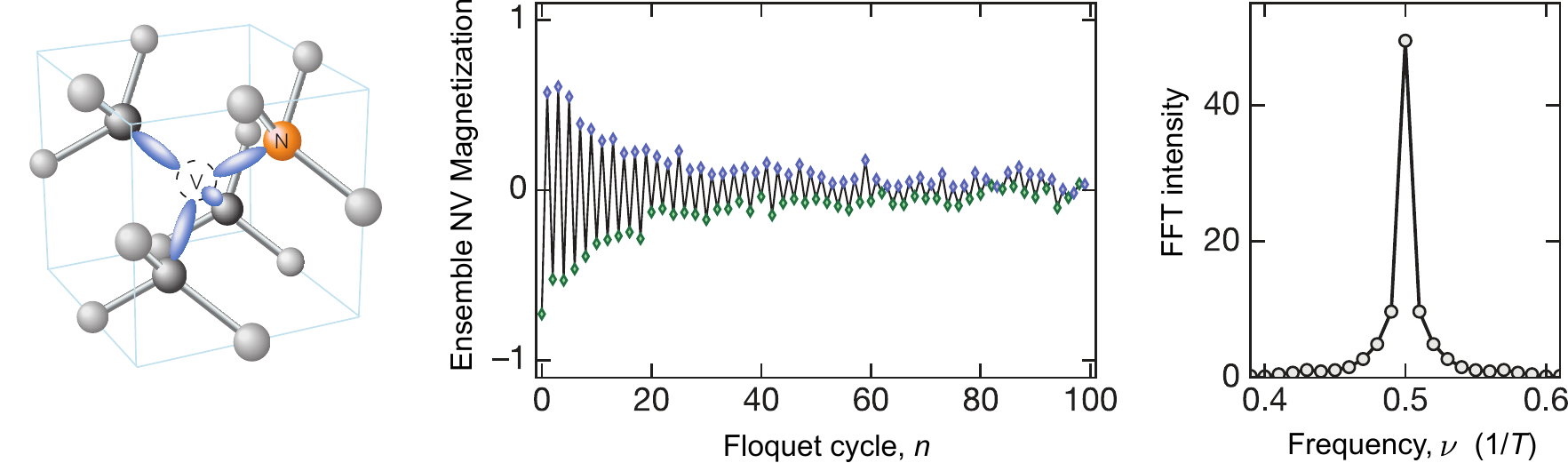}
    
    \vspace{0.3cm}
    \includegraphics[scale=0.5]{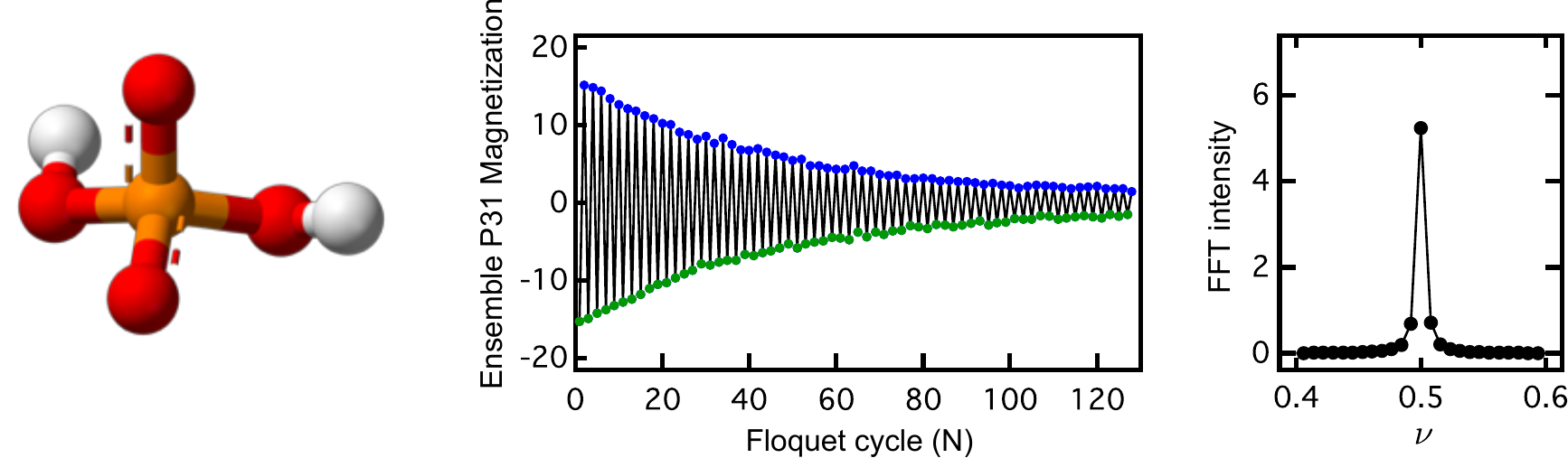}
    \caption{
    Similarities in the observed signatures of discrete time crystal behavior in three different experiments. Top row: A 1D chain of trapped ions, the oscillations in time of the magnetization of individual $^{171}$Yb$^+$ ions, and their Fourier transform. 
    Middle Row: An NV center in diamond, oscillations in the  magnetization of a dense ensemble of NV centers, and their Fourier transform. 
    Bottom row: Ammonium dihydrogen phosphate, oscillations in the bulk $^{31}$P nuclear spin magnetization, and their Fourier transform.
    In the absence of long-range interactions, perturbations to the $\pi$-pulse cause each experiment to observe a split Fourier peak corresponding to beating of the magnetization oscillations. When long-range interactions are added, the system exhibits a rigid $\nu=1/2$ subharmonic peak.
    }
    \label{Fig:three_expts}
  \end{figure}

Some of the experimental platforms and the observed signatures are summarized in Figure \ref{Fig:three_expts}.




\subsection{MBL time crystal in trapped ions}
\label{sec:trappedions}

\begin{figure}
\includegraphics[width=0.7\linewidth]{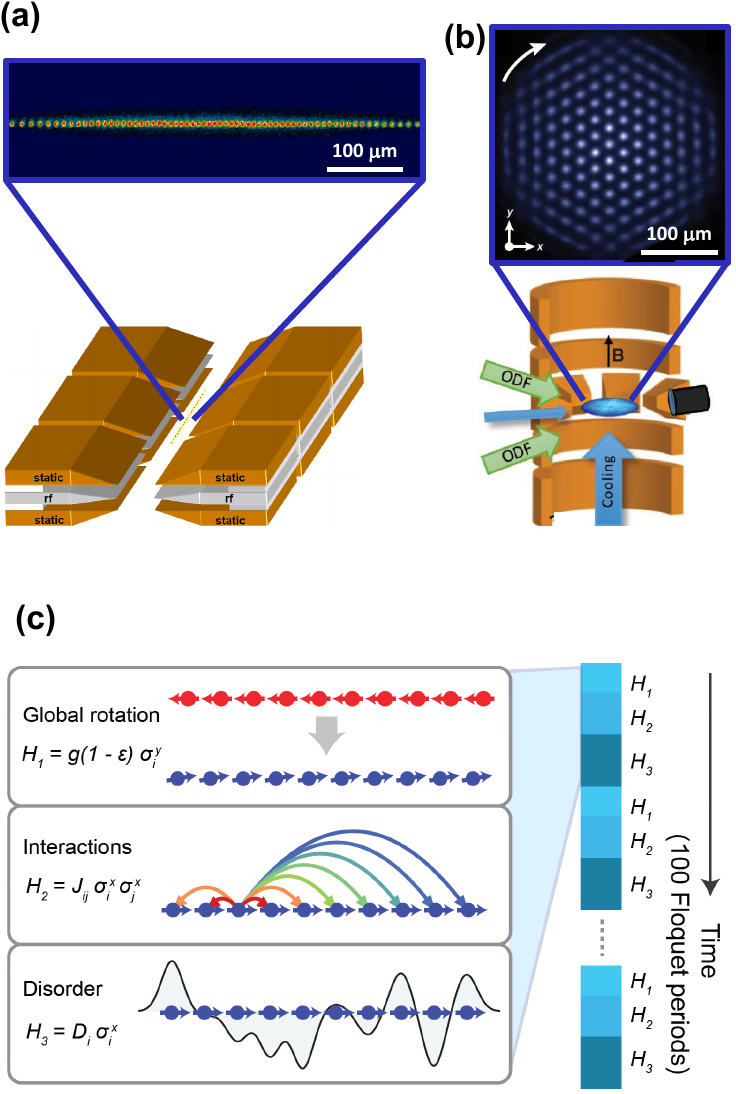}
\caption{(a) Radiofrequency (rf) linear trap used to prepare a 1D crystal of atomic ions.  For sufficiently strong transverse confinement, the ions form a linear crystal, with an image of 64 ions shown above with characteristic spacing $5$ $\mu$m for $\Yb$ ions [image taken from \cite{islam2011onset}]. (b) Penning trap used to prepare a 2D crystal of atomic ions.  For sufficiently strong axial confinement, the lowest energy configuration of the ions is a single plane triangular lattice that undergoes rigid body rotation, with an image of $\sim 200$ $^9$Be$^+$ ions shown above with a characteristic spacing of $20$ $\mu$m [image taken from \cite{bohnet2016quantum}]. (c) Schematic three-step Floquet cycle used in the trapped ion DTC experiments. The first step corresponds to global rotation (with $\pi$-pulse error $\epsilon$). The second step corresponds to strong long-range interactions. The third step corresponds to the application of controlled on-site disorder using a single-site addressing laser beam. In the case of NV centers, the Floquet protocol is similar although the disorder is not applied individually to each spin. In the case of AdP, there is negligible disorder}
\label{fig:traps}
\end{figure}

Trapped atomic ions are a versatile experimental platform for investigating time crystals with both prethermal and MBL methods \cite{ZhangJ17}. Atomic ions are confined with external electric fields, and certain ion trap geometries admit laser-cooled crystals in a variety of dimensions, as shown in Fig. \ref{fig:traps}. 

The interaction between trapped ions is inherently long-range due to the Coulomb interaction, but the relevant degree of freedom here is an internal atomic ``spin'' degree of freedom, which is represented by a pair of atomic levels behaving as an effective spin-1/2 particle or qubit. These are typically hyperfine levels that are also used as atomic frequency standards, so they enjoy fundamental $T_1$ and $T_2$ coherence times that can approach hours. The spins can be initialized through an optical pumping process: by applying resonant laser radiation that couples the spin states to appropriate short-lived excited states, each spin is initialized with $>99.9\%$ state purity in a few microseconds. The magnetization of each spin can also be measured at any time through standard spin-dependent fluorescence detection, resulting in greater than $99\%$ detection efficiency per spin.
Laser cooling can prepare the motional states of the ions to near the ground state of harmonic motion \cite{Leibfried__03}, which is important for the control of the spin-spin interactions as detailed below. 

The Coulomb interaction can be modulated with spin-dependent external classical electromagnetic forces (optical or microwave), resulting in an effective many-body Ising interaction between the spins \cite{porras2004effective}.  Accompanied by resonant spin excitations or spin level shifts that provide an effective magnetic field along any axis, it becomes possible to realize a long-range transverse-field Ising Hamiltonian of the form:
\begin{equation}
H_{ions} = \sum_{i,j} \frac{J_{ij}}{|i-j|^\alpha}\sigma_x^i\sigma_x^j + B_y\sum_{i} \sigma_y^i + \sum_{i}B^i_z \sigma_z^i.
\label{Ham}
\end{equation}
An important feature of the ion platform is that the  effective transverse $z$ field (spin level shift) can be made site-dependent, which allows for programmable disorder and is necessary for realizing MBL. 
%
Disorder in the couplings $J_{ij}$ can also be generated by controlling the detuning associated with the laser beams that modulate the spin-dependent forces \cite{korenblit2012quantum}. 
%
The Ising interaction falls off with distance as $1/r^\alpha$, where the exponent $\alpha$ can be experimentally tuned between $\alpha=0$ (infinite range) to $\alpha=3$ (dipole-dipole) by appropriately adjusting the modulating force on the atomic ions \cite{porras2004effective, deng2005effective, taylor2008wigner,islam2013emergence}.  
This tuning of the interaction range is crucial in determining the nature and effective dimensionality of the interactions.  Numerical studies suggest that the model supports MBL behavior in a 1D crystal, so long as the power-law interactions fall off sufficiently fast \cite{yao2014many,nandkishore2017many}.

To observe time crystalline behavior in the trapped ion system \cite{ZhangJ17}, the authors drive up $N=14$ atomic spins using a long-range modified version of the MBL Floquet sequence found in Eqn.~(\ref{MBLham2}), as depicted in Figure 6c \cite{Yao17, ZhangJ17}. 
For up to $\sim 10^2$ Floquet cycles, a measurement of the spin-spin autocorrelation function $\bra{\psi_0}\sigma_{i}^x(t)\sigma_{i}^x(0)\ket{\psi_0}$ is performed for each of the individual spins $i$ (Figure~\ref{Fig:three_expts}a).
In the absence of interactions, the oscillations of each spin are sensitive to the precise value of the global rotation (approximate $\pi$-pulse) induced by $H_2$ [Eq.~\ref{MBLham2}] and are therefore expected to track the perturbation $\epsilon = |g-1|$. 
This results in a splitting of the Fourier response spectrum by $2\epsilon$ in the frequency domain, precisely as expected (Figure 1b). 
Adding in the field disorder term in $H_1$ to the ion Floquet period causes the individual spins to further precess at different Larmor rates and dephase with respect to one another.
Finally by adding in the long-range Ising interactions into $H_1$ to complete the Floquet period, many-body synchronization is restored (Figure 5a) \cite{ZhangJ17}).  
The key observation here is the rigidity of the ion's  temporal response; it is locked to twice the Floquet period, even in the face of perturbations to the drive in $H_2$. 
This rigidity persists for different initial states and reasonable strengths of perturbations (depending on the effective Ising interaction strength), while for even larger perturbations the central peak amplitude decreases and the variance of the subharmonic feature increases.
Moreover, measuring this variance of the period-doubled Fourier component of the spin magnetization and plotting the maximal variance versus both the Ising interaction strength and the drive perturbation  leads to a phase diagram (Figure 1b) whose boundary is consistent with theoretical estimates \cite{Yao17}. Moreover, at the phase boundary, large fluctuations of the peak height as a function of disorder realization are observed, suggesting criticality \cite{Yao17,berdanier2018strong}.


Overall, the trapped ion system realizes arguably the cleanest textbook stroboscopic Floquet Hamiltonian for realizing a time crystal.  However, despite its cleanliness, the system sizes are quite small and certainly far from the thermodynamic limit requisite for claiming true time crystalline order.



\subsection{NV Centers in diamond}

We now turn our attention to the opposite limit and consider a truly many-body system composed of optically active spin defects in the solid-state, namely nitrogen-vacancy (NV) centers in diamond \cite{NVreview}. 
Each individual NV center constitutes an $S = 1$ electronic spin, which can be optically initialized and read out in direct analogy to the ion discussions above.
While the majority of experiments on NV centers focus on its properties as a long-lived, room-temperature quantum bit, recent experiments \cite{ChoiS17,choi2019probing} on dense NV ensembles have led to the observation of discrete time crystalline order. 
As in the ion case, three central ingredients enter the NV experiments (Figure 7). First, there exist spin level shifts of the NV resonances owing to two sources of intrinsic disorder: random positioning of the NV centers within the diamond lattice and the presence of additional paramagnetic impurities. Second, the NV centers interact with one another via long-range, magnetic dipole-dipole interactions. Third, controlled driving (to realize the requisite approximate $\pi$-pulses) can be performed via microwave excitation. In combination, this leads to the following schematic Hamiltonian:
\begin{align}
    H(t) = \sum_i\Omega_x(t)  S_i^x + \Omega_y (t)  S_i^y +\Delta_i S_i^z + \sum_{ij} (J_{ij}/r_{ij}^3) ( S_i^x S_j^x + S_i^y S_j^y - S_i^z S_j^z),
   \label{eqn:ham}
\end{align}
where, $S_i^\mu$ are Pauli spin-$1/2$ operators acting on an effective qubit spanned by two of the spin sub-levels of the NV center.
One important feature to note is that the disordered on-site  fields, $\Delta_i$, exhibit an  approximate standard deviation which is significantly larger than the average dipolar interaction strength. 
While this nominally puts the system in the ``strong'' disorder regime, as discussed in Section \ref{subsec:critical}, one does not expect localization to occur, due to the long-ranged interactions.

\begin{figure}
\includegraphics[width=0.75\linewidth]{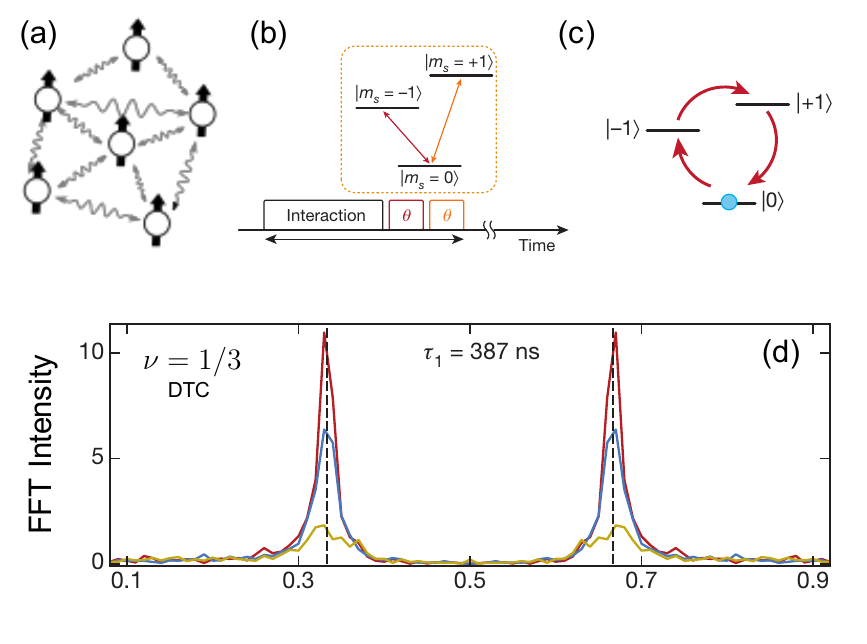}
\caption{\textbf{Experimental set-up and observation of discrete timecrystalline
order.} (a) Nitrogen Vacancy centers interact with one another via long-range dipolar interactions. (b,c) By driving all three spin sub-levels of the NV center, it is possible to implement a Floquet sequence which looks like a clock Hamiltonian that permutes the populations cyclically between the spin states. 
(d) Dipolar interactions stabilize an $n=3$ DTC within the observed number of Floquet cycles.}
\label{fig:NVdata}
\end{figure}

At its core, the central observation in the NV experiments is similar to that of the ions: interactions stabilize the sub-harmonic response of the system to perturbations of the drive (Figure 5b). 
In the NV experiments, the authors measure the ensemble magnetization of $\sim 10^6$ NV spins as opposed to the ions, where for a small 1D chain, it is possible to read out the correlation function for each ion. 
In addition to exploring $n=2$ time-crystalline order, the authors leverage the spin-1 nature of the NV center to explore the stability of an $n=3$ discrete time crystal as well. 
There, they utilize a cyclic driving scheme (Figure 7) between the NV's three sub-levels and again observe an interaction stabilized sub-harmonic Fourier peak. 
This example provides a natural intuition for how to realize an $\mathbb{Z}_n$ discrete time crystal by considering an $n$-level system and performing a clock-drive that cyclically permutes the populations. 

Perhaps the most intriguing question to ask is the following: Given the marked similarities between the observations in the ion  and  NV experiments (Figure 5), is the origin of time-crystal behavior in these two platforms the same?
Prevailing wisdom is that this is not the case. 
Unlike the 1D ion spin chain, the NV experiments are performed in a bulk three dimensional diamond sample. Coupled with the fact that the dipolar interactions exhibit a $1/r^3$ power-law tail, such a system is not expected to exhibit many-body localization despite the presence of strong disorder \footnote{Although the conventional wisdom that many-body localization can never happen with interactions that decay slower than some power has recently been questioned \cite{nandkishore2017many}.}. 
Following the flow-chart in Figure 3, one might naturally suspect a possible prethermal origin. 
However, although the NV centers are optically pumped to an extremely \emph{low entropy} state (i.e.~almost all spins in  the $|m_s =0 \rangle$ state), this state represents an extremely \emph{high temperature} state with respect to the original dipolar Hamiltonian. 
Since one only expects the existence of prethermal time crystalline order when the initial state is at low enough energy density to be ordered with respect to the prethermal Hamiltonian, it is hard to envision that such an optically-pumped NV state could lead to a prethermal time crystal. 
Rather, as explained in detail in Section \ref{subsec:critical}, the NV system may correspond to a so-called ``critical'' time crystal where the DTC order exhibits a power-law decay as the system ultimately approaches thermal equilibrium.
The experiment has not observed a clear power-law decay with time and there is evidence that the effective  disorder is in fact time-dependent, but in some other respects (for example, the phase diagram) there appears to be agreement with the predictions from a theory of critical time crystals ~\cite{ChoiS17, Ho17}; this theory, which is based upon resonance counting arguments is expected to hold only for initial transient dynamics (see Section \ref{subsec:critical}), and a deeper understanding of the late time behavior observed in the experiments requires further investigations \cite{choi2019probing}.



\subsection{NMR experiments}

Finally, we now move onto a third experimental platform which further highlights our discussion from the last section. 
Namely, that despite nominally similar observations of an interaction-stabilized, sub-harmonic response, it is possible for this response to have a distinctly different conceptual origin. 
Here, we focus on nuclear magnetic resonance experiments in ammonium dihydrogen phosphate (AdP) \cite{Rovny18a,Rovny18b}.
More broadly, the isolation associated with nuclear spins in solids makes them an interesting candidate for quantum non-equilibrium studies.
Coupled with the ability to perform coherent driving using radiofrequency fields, such systems also exhibit the necessary ingredients for realizing the types of Floquet evolution that can host time crystals \cite{Rovny18a,Rovny18b}.

The relevant degrees of freedom in  ammonium dihydrogen phosphate are the spin-1/2 $^{31}$P nuclei. Much like NV centers, these nuclear spins interact via magnetic dipolar interactions. However, the key distinction here is the complete lack of disorder. 
Unlike NVs which are randomly positioned within the diamond lattice, AdP exhibits a 100\% occupied crystal lattice. Coupled with Hahn echo measurements which suggest a small inhomogenous spread in the spin levels of the $^{31}$P nuclei, one does not expect these experiments to be dominated by disorder.
While similar analyses rule out the possibility of MBL (dipolar interactions in 3D, as with NV centers; little or no disorder, unlike with NV centers) and prethermal (high temperature initial state, as with NV centers) origins for the observed time crystalline behavior, the lack of disorder also rules out the possibility of a critical time crystal. 

In fact, the origin of the time-crystalline signature in this system is likely to be much more straightforward than any such mechanisms. In a rotating frame, the time-dependent Hamiltonian of the system nearly commutes with each single-spin magnetization $\sigma_i^z$. The ``time-crystalline'' regime observed in the experiment could then simply correspond to slow relaxation of $\langle \sigma_i^z \rangle$ in this rotating frame.
If this is the case, the decay rate of the oscillations would be determined by the strength of the integrability-breaking terms, and in particular the rate at which they induce decay of $\langle \sigma_i^z\rangle$.
Within this context, one would generally not call this phenomenon a time crystal according to the definitions of Section \ref{sec:TTSB}, because it relies on being close to integrability and is not stable to generic many-body perturbations.

Finally, we now turn to two other NMR studies of systems that are approximately described by central spins interacting with a larger number of satellite spins which don't interact with each other.
P-doped Si which has been isotopically purified so that only $0.005\%$ of the Si is $^{29}$Si, in which case the physics is approximately described by such a model: the P impurity spin interacts with the $^{29}$Si nuclear spins \cite{OSullivan18}.
There is disorder in the locations of the $^{29}$Si nuclei, but the system is 3D and has a hot initial state, as in the case of NV centers in diamond.
In the organic molecules acetonitrile, trimethylphosphite, and and tetrakis(trimethylsilyl)silane, the nuclear spins of $^{13}$C, $^{31}$P, and $^{29}$Si  interact with, respectively, 4,10, and 37 satellite H nuclear spins \cite{PalS18}. In this case, the absence of a central spin leads to oscillations that are not locked to half the frequency of the drive, but the presence of a central spin leads to oscillations at half the frequency of the drive even for quite large deviations from a perfect $\pi$-pulse. These systems are not in the thermodynamic limit, of course. Moreover, there is weak or no disorder, and the initial state is too hot to be in a prethermal DTC.

\section{Beyond time crystals}
As we have described, a time crystal is the simplest, and most readily accessible experimentally, example of a new phase of matter that occurs in isolated quantum systems. Going forward, we expect isolated quantum systems to be the setting for many more new phases of matter.

One avenue of exploration is to go beyond Floquet systems and instead consider \emph{quasiperiodically} driven systems, where the drive contains  at least two different incommensurate frequencies (that is, their ratio is an irrational number). The issue of heating is much less well understood in such systems, compared to periodically driven systems. Nevertheless, there is numerical evidence for a logarithmically slow relaxation in a quasi-periodically driven spin chain in 1-D \cite{Dumitrescu_1708}. Moreover, Ref.~\cite{Dumitrescu_1708} also found evidence for a ``time quasicrystal'' response that is a distinct regime from the ``trivially synchronized'' response. Roughly, what this means is that, whereas a quasiperiodic drive can be thought of tracing out a path at an irrational angle through a higher-dimensional periodic space, the response of the time quasi-crystal phase can be thought of as a trace through a different higher-dimensional space with a larger period.

Let us now return to Floquet systems, where in addition to spontaneous symmetry breaking phases, there can also be new \emph{topological} phases of matter. In equilibrium, although phases were once thought to be characterized purely by their pattern of spontaneous symmetry breaking, it is now known that at zero temperature there are also topological phases, which are distinguished by more subtle features of quantum entanglement, possibly in interplay with the microscopic unbroken symmetries. As we discussed in Section \ref{sec:locprotected}, in MBL systems all eigenstates have the same properties as ground states, so there can be sharply distinct regimes in such systems distinguished by the topological order of their eigenstates. However, in Floquet-MBL systems there can also be new topological phases of matter that have no static analog \cite{Khemani16,vonKeyserlingk16a,Else16a,
Po_1609,Harper_1609,Roy_1610,Po_1701,potirniche2017floquet,Haah_1812}.

Continuing the theme of eigenstate properties, one quite general way to characterize phases of matter in Floquet systems is in terms of \emph{eigenstate loops} \cite{Else16a,ElsePhd}. Consider an eigenstate $\ket{\psi}$ of the Floquet evolution operator $U_F$, and define $\ket{\psi(t)}$ to be its time evolution under the time-periodic Hamiltonian $H(t)$, i.e.
\begin{equation}
\frac{d}{dt}\ket{\psi(t)} = H(t) \ket{\psi}, \quad \ket{\psi(0)} = \ket{\psi}.
\end{equation}
Since $\ket{\psi}$ is an eigenstate of $U_F$, by definition $\ket{\psi(T)} \propto \ket{\psi}$ (up to a global phase factor). So if we mod out by the global phases, each eigenstate defines a \emph{loop} in the space of ground-state-like states. [Recall that when the system is MBL, each eigenstate $\ket{\psi}$ can be written as the ground state of some fictitious quasi-local gapped Hamiltonian. It follows that so can $\ket{\psi(t)}$ for any $t$]. One can also generalize this notion to cases with spontaneous symmetry breaking by considering an eigenstate multiplet instead of just a single eigenstate. From very general (though not completely rigorous) points of view, one can argue that the classification of such loops, where a symmetry $G$ is imposed on the instantaneous time-dependent Hamiltonian, is in one-to-one correspondence with the classification of static phases with symmetry $G \times \mathbb{Z}$. Here, the $\mathbb{Z}$ reflects the fact that Floquet systems have an inherent discrete time-translation symmetry. In a time-crystal, this symmetry is spontaneously broken, but it can give rise to non-trivial symmetry-protected phases as well.

However, eigenstates are not the whole story. There exist distinct regimes of Floquet-MBL systems that cannot be distinguished just by looking at a single eigenstate at a time \cite{Po_1609,Harper_1609,Roy_1610,Po_1701,Haah_1812}. Rather, these phases are distinguished by some non-trivial properties of the Floquet evolution operator acting on the entire Hilbert space. Finding general formalisms to describe such phases remains an open problem.

\begin{acknowledgments}
The authors gratefully acknowledge Francisco Machado and Soonwon Choi for an extremely careful reading of the manuscript and many helpful suggestions.
\end{acknowledgments}

\begin{appendices}
\section{Multiplets and the definition of spontaneous symmetry breaking}
\label{appendix:multiplets}

In Section \ref{sec:eigenstates}, we made some claims about the equivalence between different definitions of spontaneous symmetry breaking. Here, we will give a proof of these claims. We will need to make some additional technical assumptions which, however, are satisfied in all of the models that we consider.

The key assumption is that eigenstates always come in multiplets with a particular structure.
To formalize this idea, we can make the following definitions:

\begin{quote}
\textbf{Definition}. A \emph{eigenstate multiplet} of a Hamiltonian $H$ or Floquet evolution operator $U_\text{F}$ is a subspace spanned by a collection of a finite number of eigenstates of $H$ or $U_\text{F}$, such that the subspace has a basis in which each basis state (which is not necessarily an eigenstate) is short-range correlated, with the following properties:
\begin{enumerate}
\item The different short-range correlated basis states are locally distinguishable: that is, in the vicinity of any point $x$ in space, there exists a local observable $\hat{o}(x)$ whose expectation value  distinguishes between all of the different short-range correlated basis states for the multiplet. 
\item \label{src_superselection} The different short-range correlated basis states are not connectible by local operators; that is, if $\ket{\lambda}$ and $\ket{\lambda'}$ are two different short-range correlated basis states, then the matrix element
\begin{equation}
\bra{\lambda} \hat{A} \ket{\lambda'} = 0,
\end{equation}
for any local operator $\hat{A}$
(possibly up to corrections exponentially small in the system size).
\item The symmetries of $H$ or $U_F$ permute the short-range correlated basis states (possibly up to global phase factors), and the symmetry action is transitive, i.e.\ any two short-range correlated basis states are related by some symmetry.
\end{enumerate}

\end{quote}
Then we can define

\begin{quote}
\textbf{Definition}. A symmetry $u$ is \emph{spontaneously broken} in a given eigenstate multiplet if the permutation action of $u$ on the short-range correlated basis states is non-trivial, or in other words, the short-range correlated basis states are not invariant under the symmetry up to a global phase. 
\end{quote}
(For simplicity, we will only consider symmetries $u$ which are central elements in the whole symmetry group of $H$ or $U_F$, i.e.\ they commute with all the other symmetries; this is certainly true for time-translation symmetry. Combined with the transitivity condition on the permutation action, this ensures that any given short-range correlated basis state is invariant under $u$ [up to global phase] if and only if they all are.)
The physical interpretation of the short-range correlated basis states is that they are the symmetry-broken states that will emerge as non-degenerate eigenstates when all the symmetries are lifted explicitly by infinitesimal perturbations in the Hamiltonian (for example $\ket{\uparrow}$ and $\ket{\downarrow}$ in the case of a spontaneously broken Ising symmetry).

What one can show is that, in general, for a symmetry $u$, $\ket{\psi}$ a $u$-invariant eigenstate in a multiplet, and $\hat{o}(x)$ a family of local observables the expectation values of which distinguish different symmetry-breaking states in the multiplet then generically we have
\begin{equation}
\label{ssb_correlator_2}
\bra{\psi} u \hat{o}(x) u^{-1} \hat{o}(y) \ket{\psi}  - \bra{\psi} \hat{o}(x) \hat{o}(y) \ket{\psi} \nrightarrow 0 \quad \mbox{as $|x - y| \to \infty$}.
\end{equation}
even as $|x - y| \to \infty$.
Conversely, if \eqnref{ssb_correlator_2} is satisfied then $u$ must be spontaneously broken. Setting $u = U_F$ gives the unequal time correlator discussed in Section \ref{sec:defn_dtc}, whereas choosing an $\hat{o}(x)$ such that $u \hat{o}(x) u^{-1} = e^{i \alpha} \hat{o}(x)$ for some phase factor $e^{i\alpha}$ gives the off-diagonal long-range order discussed in Section \ref{sec:defn_general}.

To see \eqnref{ssb_correlator_2}, first note that, from the definition of an eigenstate multiplet, there exists a finite set $\Lambda$, and a permutation $\sigma : \Lambda \to \Lambda$, such that $\Lambda$ labels the short-range correlated basis states of the multiplet, which we write as $\{ \ket{\lambda} : \lambda \in \Lambda \}$, and $u$ acts on the basis states as $u \ket{\lambda} = \beta_\lambda \ket{\sigma(\lambda)}$, for some global phase factor $\beta_\lambda$. Let us consider a $u$-invariant (up to global phase) state $\ket{\psi}$ in the multiplet, and expand it in terms of the short-range correlated basis,
\begin{equation}
\ket{\psi} = \sum_{\lambda \in \Lambda} c_\lambda \ket{\lambda}.
\end{equation}
where the $u$-invariance of $\ket{\psi}$ implies we must have $|c_{\sigma(\lambda)}| = |c_\lambda$ for all $\lambda$.

Now we define
\begin{align}
C_\sigma(x,y) &= \sum_{\lambda} |c_\lambda|^2 o_{\sigma(\lambda)}(x) o_\lambda(y), \\
C(x,y) &= \sum_\lambda |c_\lambda|^2 o_\lambda(x) o_\lambda(y),
\end{align}
where $o_\lambda(x) = \bra{\lambda} \hat{o}(x) \ket{\lambda}$.
Clearly, if $u$ is not spontaneously broken, then $\sigma(\lambda) = \lambda$ for all $\lambda$, and hence $C_\sigma(x,y) = C(x,y)$. Moreover, if we have translational invariance such that $o_\lambda(x) = o_\lambda(y) := o_\lambda$, then we can verify that
\begin{equation}
C_\sigma(x,y) - C(x,y) = -\frac{1}{2} \sum_\lambda |c_\lambda|^2 (o_{\sigma(\lambda)} - o_\lambda)^2,
\end{equation}
and so we have conversely that $C_\sigma(x,y) = C(x,y)$ implies that the $u$ is not spontaneously broken. Without translational invariance, it is possible to have an ``accidental cancellation'' such that $C_\sigma(x,y) = C(x,y)$ even if $u$ is spontaneously broken, but certainly this would not happen generically.

Let us now show that $C_\sigma(x,y) \neq C(x,y)$ is equivalent to \eqnref{ssb_correlator_2}. Indeed, from assumption \ref{src_superselection} in the definition of eigenstate multiplet, we find that
\begin{equation}
\bra{\psi} u \hat{o}(x) u^{-1} \hat{o}(y) \ket{\psi} = \sum_{\lambda} |c_\lambda|^2 \bra{\lambda} u \hat{o}(x) u^{-1}  \hat{o}(y) \ket{\lambda},
\end{equation}
and so we see that
\begin{equation}
\label{some_other_equation}
\bra{\psi} u \hat{o}(x) u^{-1} \hat{o}(y) \ket{\psi} - C_\sigma(x,y) = \sum_\lambda |c_\lambda|^2 \mathcal{C}_{\lambda}[u\hat{o}(x)u^{-1}, \hat{o}(y)]
\end{equation}
where we have defined the connected correlator
\begin{equation}
\mathcal{C}_\lambda[\hat{A},\hat{B}] = \bra{\lambda} \hat{A} \hat{B} \ket{\lambda} - \bra{\lambda} \hat{A} \ket{\lambda} \bra{\lambda} \hat{B} \ket{\lambda}.
\end{equation}
Similarly, we find 
\begin{equation}
\label{some_equation}
\bra{\psi} \hat{o}(x) \hat{o}(y) \ket{\psi} - C(x,y) = \sum_\lambda |c_\lambda|^2 \mathcal{C}_{\lambda}[\hat{o}(x), \hat{o}(y)].
\end{equation}
As $|x-y| \to \infty$, the connected correlators on the right-hand side of Eqs.~(\ref{some_other_equation}) and (\ref{some_equation}), go to zero, and so we obtain \eqnref{ssb_correlator_2} if and only if $C_\sigma(x,y) \neq C(x,y)$.

\end{appendices}

%


\end{document}